\documentclass[reprint,aps,pre]{revtex4-1}
\usepackage{graphicx}
\usepackage{bm}
\usepackage{hyperref}
\usepackage{color}
\usepackage{amsfonts}
\usepackage{amsmath}
\usepackage{soul}

\begin{document}
\newcommand{\TODO}[1]{\textcolor{red}{#1}}
\def\NOTE#1{{\textcolor{red}{\bf [#1]}}}   
\def\DEL#1{{\textcolor{green}{\textst{#1}}}}        
\def\ADD#1{{\textcolor{blue}{#1}}}         
\def\AD#1{{\textcolor{magenta}{#1}}}       
\def\JB#1{\textcolor{red}{JB: #1}}        
\def\CH#1{{\textcolor{blue}{CH: #1}}}        
\def\itm{\par\noindent$\bullet$ }
\newcommand{\ocaaddress}{Universit\'e C\^ote
  d'Azur, CNRS, OCA, Laboratoire Lagrange, Bd.\ de l'Observatoire, Nice, France}
 
\title{Tumbling dynamics of inertial chains in extensional flow}
\author{Christophe Henry} \affiliation{\ocaaddress} \author{Giorgio
  Krstulovic} \affiliation{\ocaaddress} \author{J\'er\'emie Bec}
\affiliation{\ocaaddress} \date{\today}
\begin{abstract}
  The dynamics of elongated inertial particles in an extensional flow 
  is studied numerically by performing simulations of freely jointed 
  bead-rod chains. The coil-stretch transition and the tumbling instability
  are characterized as a function of three parameters: The Peclet number,
  the Stokes number and the chain length. Numerical results show that in 
  the limit of infinite chain length, particles are trapped in a coiled or 
  stretched state. The coil-stretch transition is also shown to depend
  non-linearly on the Stokes and Peclet number. Results also reveal that
  tumbling occurs close to the coil-stretch transition and that the 
  persistence time is a non-linear function of Stokes and Peclet numbers.
\end{abstract}
\maketitle

\section{Introduction}

The dynamics of long, flexible and deformable particles
has attracted a lot of attention in the past few years. It has notable
consequences in various applications such as: The paper-making industry
(where the dynamics of fibers directly impact the properties of paper, 
see, e.g., \cite{lundell2011fluid, cui2007flow}), DNA and polymer physics 
(where the presence of polymers in a solution can profoundly affect the 
rheology of the solution \cite{shaqfeh2005dynamics}), biological 
oceanography (with the central role played by plankton such as diatoms 
which form chain-like colonies \cite{jumars2009turbulence, 
young2012quantifying}) or atmospheric sciences (where non-spherical ice 
crystals impact particle-cloud interactions \cite{pruppacher1998microphysics}). 
The challenges associated to such complex systems include the description 
of their dynamics and conformation in complex flows as well as their
effect on rheological properties of fluids (see e.g. reviews 
\cite{shaqfeh2005dynamics,lindner2015elastic,voth2017anisotropic}).
This has led to a renewed attention on the dynamics of complex-shaped 
objects such as solid spheroids \cite{pumir2011orientation,parsa2012rotation,
vincenzi2013orientation,gustavsson2014tumbling,gupta2014elliptical}, 
ellipsoids \cite{chevillard2013orientation}, solid helicoids 
\cite{gustavsson2016preferential} as well as flexible objects such as
elastic dumbbells \cite{musacchio2011deformation} or trumbbells 
\cite{plan2016tumbling,ali2016semiflexible}, flexible fibers 
\cite{brouzet2014flexible,perkins1994direct,perkins1995stretching,verhille20163d}. 

In this paper, we focus on the dynamics of elongated and deformable
particles. Recent experimental studies on elastic filaments have
revealed a complex non-linear dynamics characterized by coil-stretch
transitions (i.e. the shift from extended to folded conformations),
tumbling and buckling instabilities in various simple flows (such as
extensional flows \cite{kantsler2012fluctuations,hsiao2017direct},
shear flows \cite{harasim2013direct}) as well as more complex flows
(such as random flows \cite{liu2010stretching}, chaotic flows
\cite{chertkov2000polymer}, the flow past an obstacle
\cite{lopez2015deformation} or in a microchannel
\cite{strelnikova2017direct}). The ability of such elongated particles
to deform under various flow conditions has also been characterized in
numerous theoretical and numerical studies (see
e.g. \cite{cruz2012review,
  larson2005rheology,lindner2015elastic,shaqfeh2005dynamics,underhill2004coarse}
and references therein). In particular, the coil-stretch transition
has been reproduced using various level of descriptions for elongated
particles, including fine simulations based on the slender-body theory
(SBT) \cite{lindner2015elastic, kratky1949rontgenuntersuchung} or
coarse-grained models such as the freely jointed bead-rod, bead-spring
models or even simple dumbbells \cite{larson2005rheology,
  bird1987dynamics2}. The coil-stretch (CS) transition and tumbling
statistics have been characterized recently using simple
representations of elongated particles such as rigid rods/spheroids
\cite{marchioli2010orientation,mortensen2008dynamics}, spring
dumbbells~\cite{puliafito2005numerical} or
trumbbells~\cite{ali2016semiflexible, plan2016tumbling}). For
instance, the phenomenon of tumbling, which is associated to elongated
particles in shear flows, has been shown to occur even in stretching-
dominated flow using a trumbbell model immersed in a planar
extensional flow \cite{plan2016tumbling}.

Following these recent studies, the aim of the present study is to
study the CS transition and tumbling instability in the case of
infinitely long particles in extensional flows with stochastic
noise. For that purpose, we focus on freely jointed bead-rod chains
where fibers are discretized as a chain of elementary beads connected
by rigid rods (also called Kramers chains
\cite{kramers1944viscosity}). We introduce a high-order numerical
method to impose the holonomic constraints in the presence of a
stochastic term. A similar approach based on bead-spring chains has
recently shown that fibers are trapped in either coiled or stretched
states in the limit of infinite chain sizes and small-enough diffusion
\cite{beck2007ergodicity} in linear and non-linear extensional
flows. It remains to be seen whether such conclusions remain valid
when inertial effects are taken into account and when a bead-rod
description of long particles is used. This is the scope of the
present article.

For that purpose, the dynamics of inertial fibers in a flow is
presented in Section~\ref{sec:model_theory} which provides details on
the model used for bead-rod chains (see Section~\ref{sec:eq_motion})
together with a theoretical analysis of the stationary states (in
Section ~\ref{sec:stat_state}) as well as details on the numerical
implementation used (see Section~\ref{sec:implement}). Then, numerical
results are analyzed first for the Coil-Stretch transition in
Section~\ref{sec:CST} and then for fiber tumbling in
Section~\ref{sec:tumbling}.

\section{Model and theoretical analysis}
 \label{sec:model_theory}

\subsection{Dynamics of inertial fibers}
 \label{sec:eq_motion}

 \subsubsection{Equation of motion for inertial fibers}
  \label{sec:eq_motion_inertial}

\begin{figure}[ht]
  \begin{center}
    \includegraphics[width=.66\columnwidth]{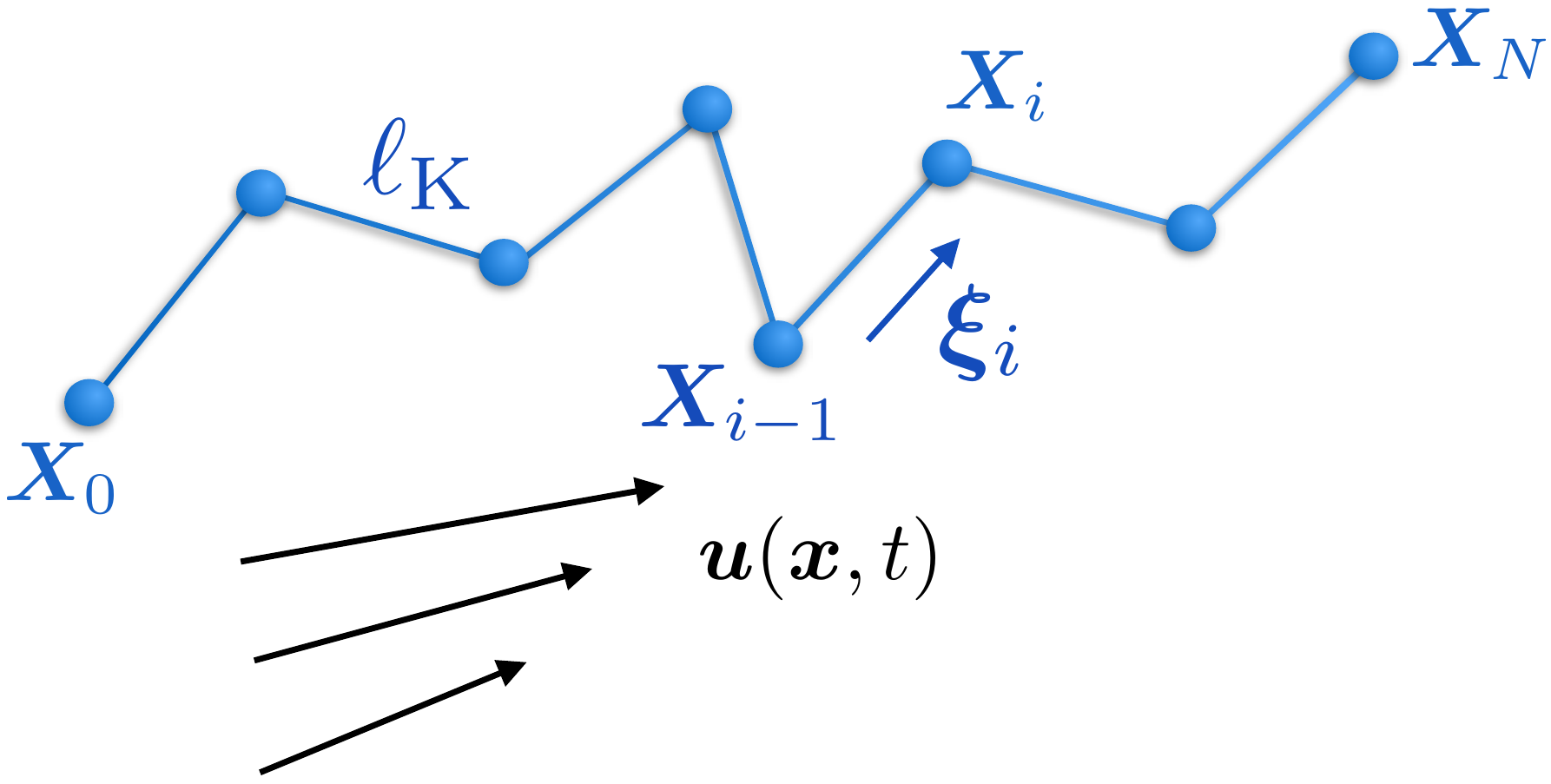}
    \caption{\label{fig:sketch} The flexible fiber is approximated as
      a bead-rod Kramers chain. The beads $\bm X_i$'s are linked
      together by infinitesimal rigid rods of length $\ell_{\rm
        K}$. The orientation of the links is given by the unitary
      vectors $\bm\xi_{i} = (\bm X_i-\bm X_{i-1})/\ell_{\rm K}$.}
  \end{center}
\end{figure}

\paragraph{Generic case.}
We consider a suspension of inertial fibers embedded in an ambient
flow and experiencing a viscous drag. Each fiber is represented as a
bead-rod Kramers chain, i.e. constituted of $N+1$ beads connected by
$N$ rigid bonds . Each bead labeled $i$ has a given position denoted
by $\bm X_i$ and is maintained at a fixed distance $\ell_{\rm K}$ from
consecutive beads in the chain (corresponding to the Kuhn
length). Each bead can undergo a free angular motion relative to its
neighbors while remaining at a fixed distance $\ell_{\rm K}$ from them
(see Fig.~\ref{fig:sketch}).  We have further assumed that all beads
are identical with a given mass $m$.
Following Newton's second law, the individual motion of each bead reads
\begin{eqnarray}
  m\, \frac{{\rm d}^2 \bm X_i}{{\rm d} t^2} &=&
  -\zeta\left[\frac{{\rm d} \bm X_i}{{\rm d} t} 
	      -\bm u(\bm X_i,t)\right]
  +\sqrt{2k_{\rm B}T\, \zeta}\,\bm\eta_i \nonumber\\
  && +\lambda_{i}\,(\bm X_i - \bm X_{i-1}) - \lambda_{i+1}\,(\bm X_{i+1}-\bm X_i),
     \label{eq:evol_X}
\end{eqnarray}
The first term on the right-hand side (RHS) corresponds to the Stokes
drag of the beads with a prescribed velocity field $\bm u$ and $\zeta$
denotes the individual drag coefficient of the particles. The second
term on the RHS stands for the effect of thermal fluctuations on each
bead: $\bm\eta_i$ are independent isotropic white noises, with $T$
denoting the fluid absolute temperature and $k_{\rm B}$ the Boltzmann
constant.  The third and fourth terms on the RHS accounts for the
tension (or internal forces) within a fiber: $\lambda_i$ is thus the
tension between the $i$-th and the $(i-1)$-th beads. Unlike
spring-bead models where the tension is given by a spring force (such
as Hookean spring as in \cite{rouse1953theory}), the tensions are here
time-dependent Lagrange multipliers associated to the the holonomic
constraint $|\bm X_i-\bm X_{i-1}| = \ell_{\rm K}$, which implies that
\begin{eqnarray}
\frac{{\rm d}}{{\rm d}t}\left|\bm X_{i}-\bm
  X_{i-1}\right|^2&=&\frac{{\rm d}^2}{{\rm d}t^2}\left|\bm X_{i}-\bm
                      X_{i-1}\right|^2=0.
\end{eqnarray}
Using Eq.\eqref{eq:evol_X}, the tensions satisfy the system
\begin{eqnarray}
 0 &=& \left[\bm X_{i}-\bm X_{i-1}\right] \cdot \left\{ \zeta \left[\bm u(\bm X_i,t) - \bm u(\bm X_{i-1},t) \right] \right. \nonumber \\ 
   && + \sqrt{2k_{\rm B}T\, \zeta}\,\left[\bm\eta_i - \bm\eta_{i-1}\right]  + 2 \lambda_{i}\,(\bm X_i - \bm X_{i-1}) \nonumber\\
   && - \left. \lambda_{i+1}\,(\bm X_{i+1}-\bm X_i) - \lambda_{i-1}\,(\bm X_{i-1}-\bm X_{i-2}) \right\} \nonumber \\
   && + \left|\frac{{\rm d}(\bm X_{i}-\bm X_{i-1})}{{\rm d}t}\right|^2.
 \label{eq:lambda}
\end{eqnarray}

\paragraph{Small fibers in an extensional flow.}
In the following, we consider that the whole fiber size is below the
smallest scale of variation of the fluid velocity field $\bm u$. In that
case, the flow stretching is uniform along the chain since all beads 
feel the same value $\bm\nabla\bm u$ of the fluid gradient. It is then
more natural to reformulate the dynamics in terms of the link unitary
directions $\bm\xi_{i} = (\bm X_i-\bm X_{i-1})/\ell_{\rm K}$, that is
\begin{eqnarray}
 \frac{{\rm d}^2 \bm \xi_{i}}{{\rm d} t^2} &=&
  -\frac{\zeta}{m}\left[\frac{{\rm d} \bm\xi_{i}}{{\rm d} t}
                                               -\bm\xi_i\cdot\bm\nabla\bm
                                               u\right]
+2\lambda_i\,\bm \xi_i - \lambda_{i+1}\,\bm \xi_{i+1}\nonumber\\
  &&- \lambda_{i-1}\,\bm \xi_{i-1}+\sqrt{\frac{2k_{\rm
     B}T\, \zeta}{m^2\ell_{\rm K}^2}}\,(\bm\eta_i-\bm\eta_{i-1}),
     \label{eq:evol_Xi}
\end{eqnarray}
The $\lambda_i$'s are again Lagrangian multipliers associated to the
holonomic constraint (constant distance $\ell_{\rm K}$ between beads), 
this time reading $|\bm\xi_i|=1$.

We further assume that the fluid gradient is given by the simple case of 
a 2D extensional flow: The fluid is stretching in the horizontal direction 
$x$ while it is compressing in the vertical direction $y$ (see also 
Fig.~\ref{fig:sketch}). The velocity gradient then simplifies to:
\begin{equation}
 \bm\nabla\bm u = \begin{pmatrix} \sigma & 0 \\ 0 & -\sigma \end{pmatrix},
\end{equation}
where $\sigma>0$ denotes the local fluid velocity shear rate. 
 \subsubsection{Dimensionless parameters}
  \label{sec:dimension_param}

  With the assumption that $\bm\nabla\bm u$ depends on a single time
  scale $\tau_{\rm fluid} = \sigma^{-1}$, the problem depends upon
  three dimensionless parameters:
\begin{itemize}
\item The \textbf{Stokes number}, which is the ratio between the beads
  response time and the fluid flow timescale
  \begin{equation}
    {\rm St} = \frac{\tau_{\rm beads}}{\tau_{\rm fluid}} = \frac{\sigma\,m}{\zeta};
  \end{equation}
  It measures the inertia of fibers: $\rm St\ll 1$ corresponds to the
  case of tracers (i.e. particles following the streamlines) while
  $\rm St\gg1$ designate particles that depart from the fluid
  streamlines.
\item The \textbf{P\'eclet number}, given by the ratio between the
  diffusion and the advection timescales
  \begin{equation}
    {\rm Pe} = \frac{\tau_{\rm diff}}{\tau_{\rm fluid}} =
    \frac{\sigma\,\ell_{\rm K}^2\,\zeta}{k_{\rm B}T};
  \end{equation}
  It measures the relative importance between thermal fluctuations and
  fluid stretching: $\rm Pe\ll1$ signify that thermal fluctuations
  dominate the dynamics while $\rm Pe\gg1$ imply that the dynamics is
  governed by fluid stretching.
\item The \textbf{number of Kuhn links} forming the chain
  \begin{equation}
    N = \frac{L}{\ell_{\rm K}}
  \end{equation}
  where $L$ denotes the total length of the chain fiber.  It is a
  measure of the number of degree of freedom in the chain.
\end{itemize}

 \subsubsection{Overdamped limit}
  \label{sec:eq_motion_overdamped}

When the relaxation time of a bead $\tau_{\rm beads}=m/\zeta$ is much shorter than the fluid
flow timescale, the inertial term in Eq~(\ref{eq:evol_Xi}) can be neglected. In this 
overdamped case, the equation simplifies to:
\begin{eqnarray}
 \frac{{\rm d} \bm\xi_{i}}{{\rm d} t} & = & \bm\xi_i\cdot\bm\nabla\bm u +  
 \sqrt{\frac{2k_{\rm B}T}{\ell_{\rm K}^2\,\zeta}}\,(\bm\eta_i-\bm\eta_{i-1})\nonumber\\
 && + 2\lambda_i'\,\bm \xi_i - \lambda_{i+1}'\,\bm \xi_{i+1}
 - \lambda_{i-1}'\,\bm \xi_{i-1},
     \label{eq:evol_Xi_tracers}
\end{eqnarray}
with $\lambda'$ the renormalized Lagrange multipliers given by
$\lambda' = \lambda\,{m}/{\zeta}$. As in the inertial case, the tension
forces are calculated by imposing a constant distance between
consecutive beads which gives the following matrix equation:
\begin{equation}
 \bm\xi_i \, \cdotp \, \frac{{\rm d}\bm\xi_i}{{\rm d}t} = 0
 \label{eq:Lambda_tracers}
\end{equation}

In presence of stochastic forces, both systems (Eqs \eqref{eq:lambda}
and \eqref{eq:Lambda_tracers}), need to be solved using high order
methods to avoid systematic numerical errors on the distance between
beads (details on the numerical implementation are provided in
Sec.~\ref{sec:implement} and in the Appendix).

\subsubsection{Observables}
\label{sec:eq_motion_observable}

Since we are interested in the stretching and orientation of such elongated fibers,
two observables have been retained to monitor their dynamics in an extensional flow
based on the analogy with spin systems:
\begin{itemize}
 \item The first quantity is the fiber coarse extension along the stretching 
 direction, which is defined as
\begin{equation}
\mathcal{L}(t)  = \frac{1}{N}\sum_{i=1}^N  s_i,
\end{equation}
with $s_i(t) = \mbox{sign}\,\xi_i^x(t)$. This quantity is analogous to 
magnetization in spin systems.
 \item The second quantity is the fiber coarse orientation along the stretching 
 direction, which is defined as
\begin{equation}
  \mathcal{F}(t)  = \frac{1}{N-1}\sum_{i=1}^{N-1}  s_i\,s_{i+1},
\end{equation}
It measures the relative orientation of consecutive beads in the fiber and is
analogous to the magnetic energy in spin systems.
\end{itemize}

In the following, we characterize the evolution of fibers in terms of
theses two quantities and as a function of the three parameters of the
system: the number of links $N$, the Peclet number $\rm Pe$ and the
Stokes number $\rm St$.

\subsection{Stationary states}
\label{sec:stat_state}

In the absence of noise, all configurations where the unitary link
vectors $\bm\xi_i$ are aligned with the stretching direction $x$ are
steady solutions to Eq.~(\ref{eq:evol_Xi}).  Indeed, if we assume that
$\bm\xi_i = (\varepsilon_i,0)^\mathsf{T}$ with
$\varepsilon_i = \pm 1$, the dynamics is trivially stationary if the
tensions satisfy
\begin{equation}
  -2\varepsilon_{i}\,\lambda_{i}+\varepsilon_{i+1}\,\lambda_{i+1}
  +\varepsilon_{i-1}\,\lambda_{i-1} =
  \frac{\zeta\,\sigma}{m}\,\varepsilon_i,
  \label{eq:steady_lambda}
\end{equation}
This system always admits solutions of the form
$(\lambda_1,\dots,\lambda_N)^\mathsf{T} =
(\zeta\,\sigma/m)\,\mathbb{E}^{-1}\,
(\varepsilon_1,\dots,\varepsilon_N)^\mathsf{T}$, where $\mathbb{E}$
denotes the tridiagonal matrix with elements
$\mathbb{E}_{i,j} = \varepsilon_{i+1}\delta_{i+1,j} -2
\varepsilon_{i}\delta_{i,j} +\varepsilon_{i-1}\delta_{i-1,j}$ and
whose determinant reads
$\det \mathbb{E} =
(-1)^N\,(N+1)\,\varepsilon_1\cdots\varepsilon_N\neq0$.  These
stationary configurations comprise the case of a fully stretched chain
for which $\varepsilon_i = +1$ for all $i$, as well as the alternating
folded polymer associated to $\varepsilon_i = (-1)^i$.  Besides these
extremes, other intermediate configurations are allowed corresponding
to partial folding of the chain.

If such stationary configurations are stable, they could play an
important role in the dynamics. For instance, the system could become
meta-stable and in presence of noise, spend some time in those states.
To evaluate the linear stability of these various cases, let us
consider that the link vectors are of the form
$\bm\xi_i = (\varepsilon_i\,\sqrt{1-\alpha_i^2},\alpha_i)^\mathsf{T}$
with $\alpha_i\ll 1$. Clearly, the perturbation of the stationary
state is order $\alpha^2$ in the $x$ direction, as well as for the
tension equation (\ref{eq:steady_lambda}). The dominant evolution is
thus in the $y$ direction and reads
$$ \frac{\mathrm{d}^2\alpha_i}{\mathrm{d}t^2} = -\frac{\zeta}{m}\left[
  \frac{\mathrm{d}\alpha_i}{\mathrm{d}t} + \sigma\,\alpha_i\right]
+2 \lambda_i\alpha_i -\lambda_{i+1}\alpha_{i+1}
-\lambda_{i-1}\alpha_{i-1},
$$
which can be written in vectorial form as
\begin{eqnarray}
  &&\frac{\mathrm{d}}{\mathrm{d}t}
     \left(\begin{array}{c} \bm\alpha\\
             \dot{\bm\alpha} \end{array}\right)
  = \mathcal{M}\, \left(\begin{array}{c} \bm\alpha\\
                          \dot{\bm\alpha} \end{array}\right),
  \nonumber \\
  &&\mbox{with } \mathcal{M} = \left(\begin{array}{cc} \mathbb{O}_N & \mathbb{I}_N\\
                                       -[(\zeta\,\sigma/m)\,\mathbb{I}_N+\bm\Delta_\lambda]
                                                                    &-(\zeta/m)\, \mathbb{I}_N
              \end{array}\right)\!.
\end{eqnarray}
$\mathbb{O}_N$ and $\mathbb{I}_N$ denote the $N\times N$ zero and
identity matrices, respectively. We have introduced
$\bm\alpha = (\alpha_1,\dots,\alpha_N)^\mathsf{T}$,
$\dot{\bm\alpha} = (\mathrm{d}\alpha_1/\mathrm{d}t,\dots,
\mathrm{d}\alpha_N/\mathrm{d}t)^\mathsf{T}$, and $\bm\Delta_\lambda$ the
tridiagonal matrix with elements
$(\Delta_\lambda)_{i,j} = \lambda_{i+1}\delta_{i+1,j} -2
\lambda_{i}\delta_{i,j} +\lambda_{i-1}\delta_{i-1,j}$. 

The linear stability of stationary configurations is entailed in the
eigenvalue $\mu_\mathcal{M}$ of the matrix $\mathcal{M}$ which has the
largest real part.  One can easily check that for all stationary
configuration obeying (\ref{eq:steady_lambda}), the vector
$\bm\alpha = (\varepsilon_1,\dots,\varepsilon_N)^\mathsf{T}$ and
$\dot{\bm\alpha} = \mu\,\bm\alpha$ is an eigenvector associated to the
eigenvalue $\mu$ if the later satisfies
$ \mu^2 + (\zeta/m)\,\mu + 2\,\zeta\,\sigma/m = 0$. We thus obtain the
following lower bound
\begin{equation}
  \mbox{Real}\,\mu_{\mathcal{M}} \ge \frac{\sigma}{2\,{\rm St}} \mbox{Real}\left[
    \sqrt{1-8\,{\rm St}}-1\right].
  \label{eq:lower_bound_muM}
\end{equation}
The right-hand side has a non-monotonic behavior as a function of the
Stokes number $St$. We can thus expect some stationary states to get
stabilized by a moderate inertia.  However these states necessarily
become less stable when $\rm St\to\infty$. This is illustrated below.

\subsubsection{The stretched line}
\label{sec:stat_state_stretched}
\begin{figure}[ht]
  \includegraphics[width=\columnwidth]{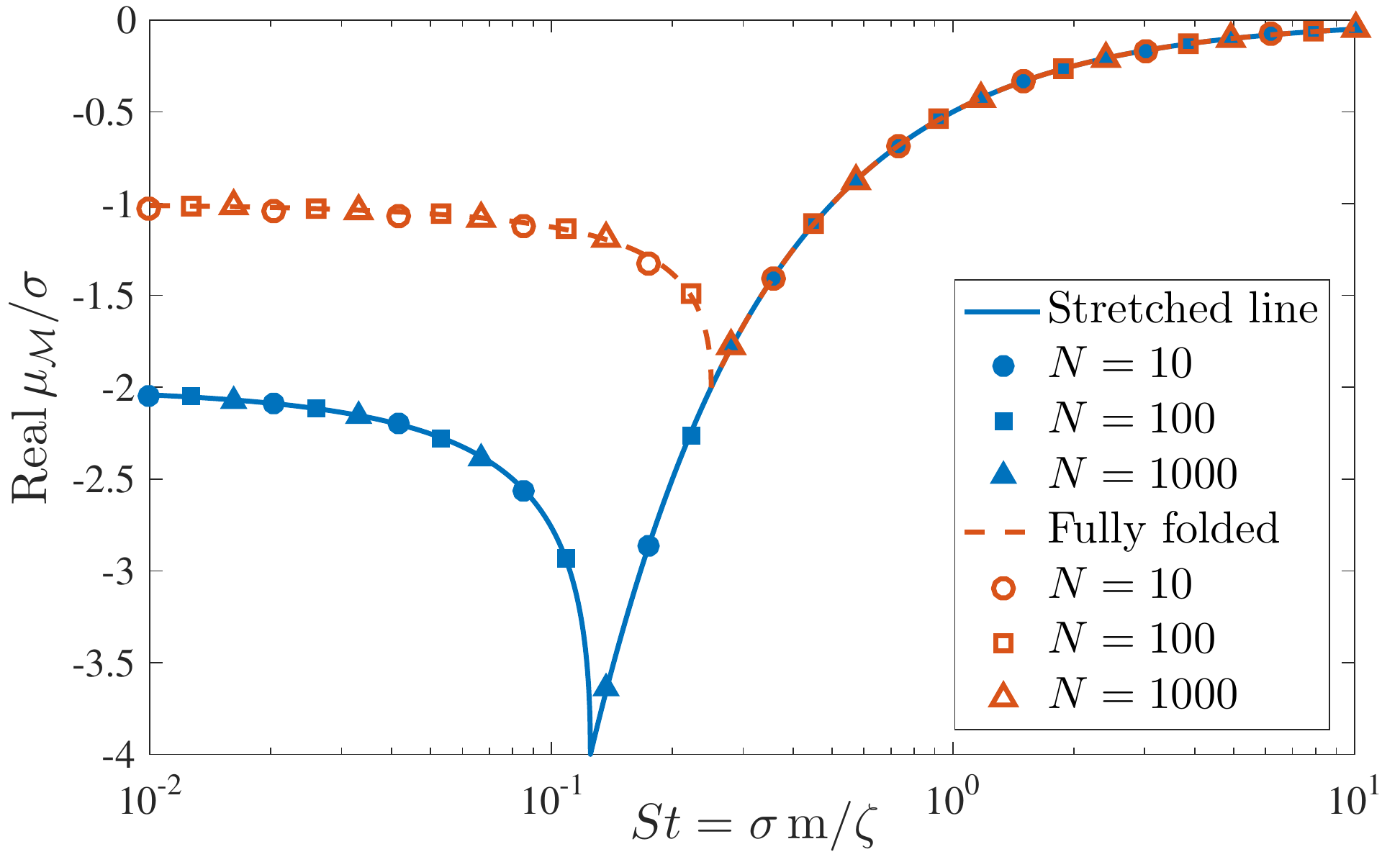}
  \caption{Real part of the most unstable eigenvalue
    $\mu_{\mathcal{M}}$ (in units of the fluid shear rate $\sigma$) as
    a function of the Stokes number $St$ for the fully stretched chain
    (solid line, filled symbol) and the completely folded case (dashed
    line, empty symbols). The lines show the predictions (see text),
    while symbols are numerical evaluations for various chain lengths,
    as labeled.}
  \label{fig: eigval_stretch_fold}
\end{figure}

Let us first consider the stationary state where all rods are aligned
with the stretching direction, i.e.\ $\varepsilon_i= \varepsilon$ for
all $i$, with $\varepsilon=\pm 1$. Equation~(\ref{eq:steady_lambda})
becomes
$$
-2\lambda_{i}+\lambda_{i+1}+\lambda_{i-1} =
{\zeta\,\sigma}/{m}\quad\mbox{with } \lambda_{0}= \lambda_{N+1}=0,
$$
so that the tensions read
\begin{equation}
  \lambda_i = -(\zeta\,\sigma/(2\,m))\,i\,(N+1-i).
\end{equation}
As can be seen from Fig.~\ref{fig: eigval_stretch_fold}, this
configuration is always stable, independently of the Stokes number and
the number $N$ of Kuhn lengths in the chain. In addition, the less
stable eigenvalue $\mu_{\mathrm{M}}$ is exactly equal in that case to
the lower bound (\ref{eq:lower_bound_muM}) discussed above. Inertia
tends to stabilize this configuration up to ${\rm St} = 1/8$. Above
this value, the stretched line become less stable with
$\mbox{Real}\,\mu_{\mathcal{M}} = \sigma/(2\,{\rm St})$.

\subsubsection{Folded polymer}
\label{sec:stat_state_folded}
A second stationary configuration of interest is the case when the
chain is completely folded in an accordion shape. We have in that case
$\varepsilon_i = (-1)^i$, so that the equations for the tensions
become
\begin{equation} 
  2\lambda_i+\lambda_{i+1}+\lambda_{i-1} = -{\zeta\,\sigma}/{m},
\end{equation}
which yields
\begin{equation}
  \lambda_i = -\frac{\zeta\,\sigma}{4\,m}\left [1-
    (-1)^i\right]-\frac{\zeta\,\sigma}{4\,m}\frac{i\,(-1)^i}{N+1}[1+(-1)^N].
  \label{eq:tension_accordion}
\end{equation}
This time, the associated matrix $\bm\Delta_\lambda$ admits zero
modes. Indeed, if without loss of generality we assume that $N$ is
odd, the second-term in the right-hand side of
(\ref{eq:tension_accordion}) vanishes and the $\lambda_i$ alternate
between $-\zeta\,\sigma/(2\,m)$ and 0. Any vector $\bm\alpha$ with
vanishing odd components belongs to the kernel of $\bm\Delta_\lambda$. As a
consequence, vectors of the form $(\bm\alpha,\mu\,\bm\alpha)$ are
eigenvectors of $\mathcal{M}$ associated to the eigenvalue
$$
\mu = \frac{\sigma}{2\,{\rm St}} \left[ \sqrt{1-4\,{\rm
      St}}-1\right].
$$
As can be seen in Fig.~\ref{fig: eigval_stretch_fold}, such
eigenvalues correspond to the less stable mode of this stationary
configuration. As for the stretched line, a sufficiently small inertia
has a stabilizing effect. The critical value is this time $1/4$, so
that when ${\rm St} >1/4$, the configuration in accordion shape is as
stable as the fully stretched chain.

\subsubsection{Intermediate configurations}
\label{sec:stat_state_inter}
\begin{figure}[ht]
  \includegraphics[width=\columnwidth]{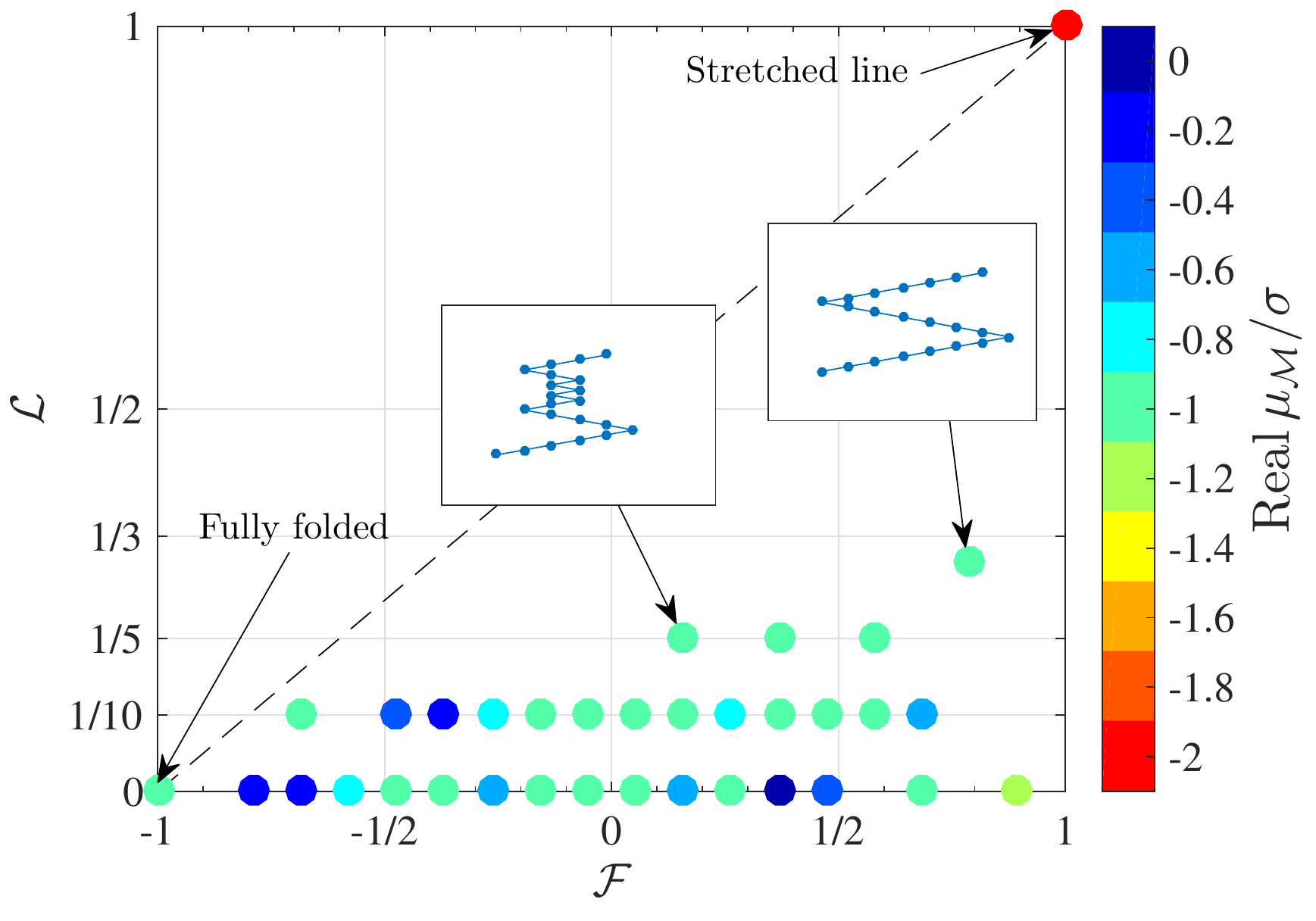}
  \caption{Stable configurations of a chain with $N=20$ and
    ${\rm St}=0$ in the $(\mathcal{F},\mathcal{L})$ plane. The filled
    circles stand for the real part of the less stable eigenvalue
    $\mu_{\mathcal{M}}$ (in units of $\sigma$). Two intermediate
    stable configurations are singled out at $\mathcal{L}\approx1/3$,
    $\mathcal{F}=(N-5)/(N-1)$, where the fiber is approximately folded
    in three equal pieces, and at $\mathcal{L}=1/5$ and
    $\mathcal{F}\approx 0.15$, for which the fiber alternates between
    stretched and folded segments.}
  \label{fig:eigval_diffstates}
\end{figure}

Besides these two extreme configurations, there exists in general a
large number of possible stationary states, as illustrated in
Fig.~\ref{fig:eigval_diffstates} in the over-damped (${\rm St}=0$)
case for $N=20$. These various configurations where obtained
numerically by a Monte--Carlo method.  The fully stretched line (in
the top-right corner) is in that case the most stable configuration
with $\mbox{Real}\,\mu_{\mathcal{M}} = -2\,\sigma$. The fully-folded
accordion shape in the bottom-left corner is associated to
$\mbox{Real}\,\mu_{\mathcal{M}} = -\sigma$. Other stable
configurations associated to partial foldings of the chain span the
bottom half of the $(\mathcal{F},\mathcal{L})$ plane.  We expect the
fiber to explore them in presence of noise. This will be the case
during the tumbling of the chain, as we will see later in
Sec.~\ref{sec:tumbling}.

\subsection{Numerical implementation}
\label{sec:implement}

In the presence of noise, we simulate the dynamics of fibers by
integrating numerically their equation of motion given by
Eq.~(\ref{eq:evol_Xi}). For that purpose, we resort to an explicit
first-order Euler--Maruyama method with temporal discretization.  In
that case, the discretized system reads:
\begin{eqnarray}
  \bm V_{i} (t+\Delta t) & = & \bm V_i(t) + \Delta t
                         \left\{-(\zeta/m)\left[\bm V_{i}(t)
                         -\bm\xi_i(t)\cdot\bm\nabla\bm
                         u\right] \right. \nonumber \\
                   && \left.+ 2\lambda_i\,\bm \xi_i(t) - \lambda_{i+1}\,\bm \xi_{i+1}(t)
                      - \lambda_{i-1}\,\bm \xi_{i-1}(t) \right\}\nonumber \\
                   && +\sqrt{\frac{2k_{\rm
                      B}T\, \zeta}{m^2\ell_{\rm K}^2}}\,(\Delta \bm
                      W_i- \Delta\bm W_{i-1}), \nonumber \\
  \bm \xi_{i}(t+\Delta t) & = & \bm \xi_{i}(t) + \Delta t\,\bm V_{i}(t),
     \label{eq:evol_Xi_num}
\end{eqnarray}
where $\bm \xi_{i}$ is the segment labeled '$i$' and
$\bm V_{i} = {\rm d}\bm \xi_{i}/{\rm d}t$ its velocity, $\Delta t$ is
the time step used in the simulation and $\Delta \bm W_{i}$ are the
increment of a two-dimensional Wiener process over a time step
$\Delta t$.

To close the system given by Eq.~(\ref{eq:evol_Xi_num}), the holonomic
constraint $|\bm\xi_i(t+\Delta t)|^2=|\bm\xi_i(t)|^2$ is used to
evaluate the tensions $\lambda_i$.  Replacing with the discretized
system, this constraint reads:
\begin{equation}
 	2\bm\xi_i(t) \, \cdotp \,   \, \bm V_i(t)
        +\Delta t \left| \bm V_i(t)\right|^2 = 0.
 \label{eq:lambda_impl}
\end{equation}
This matrix equation depends non-linearly on $\bm\xi_i(t)$ and
$\lambda_{i}$. As a result, there is no easy numerical resolution of
the tensions $\lambda_{i}$. However, we note that tension forces have
the same scaling as the stochastic term and it can be written as a
series expansion in powers of $(\Delta t)^{1/2}$.  This provides a new
high-order method for the numerical simulation of bead-rod chains
(further details are given in the Appendix).

\section{Coiled-stretched transition}
\label{sec:CST}

\subsection{Principle and mechanism}
\label{sec:CST_princ}
Fiber orientation changes constantly due to the competition between
fluid stretching and thermal fluctuations. As a result, fibers explore
a wide range of states and this leads to a broad distribution in the
fiber length and orientation with time. In particular, a fiber can
reach two notable states that are displayed in
Fig.~\ref{fig:two_states_CS_FL} for the over-damped case
${\rm St} = 0$ (together with the probability density function of the
fiber extension $\mathcal{L}(t)$ and orientation $\mathcal{F}(t)$):
\begin{figure*}[htbp]
\includegraphics[width=.48\textwidth]{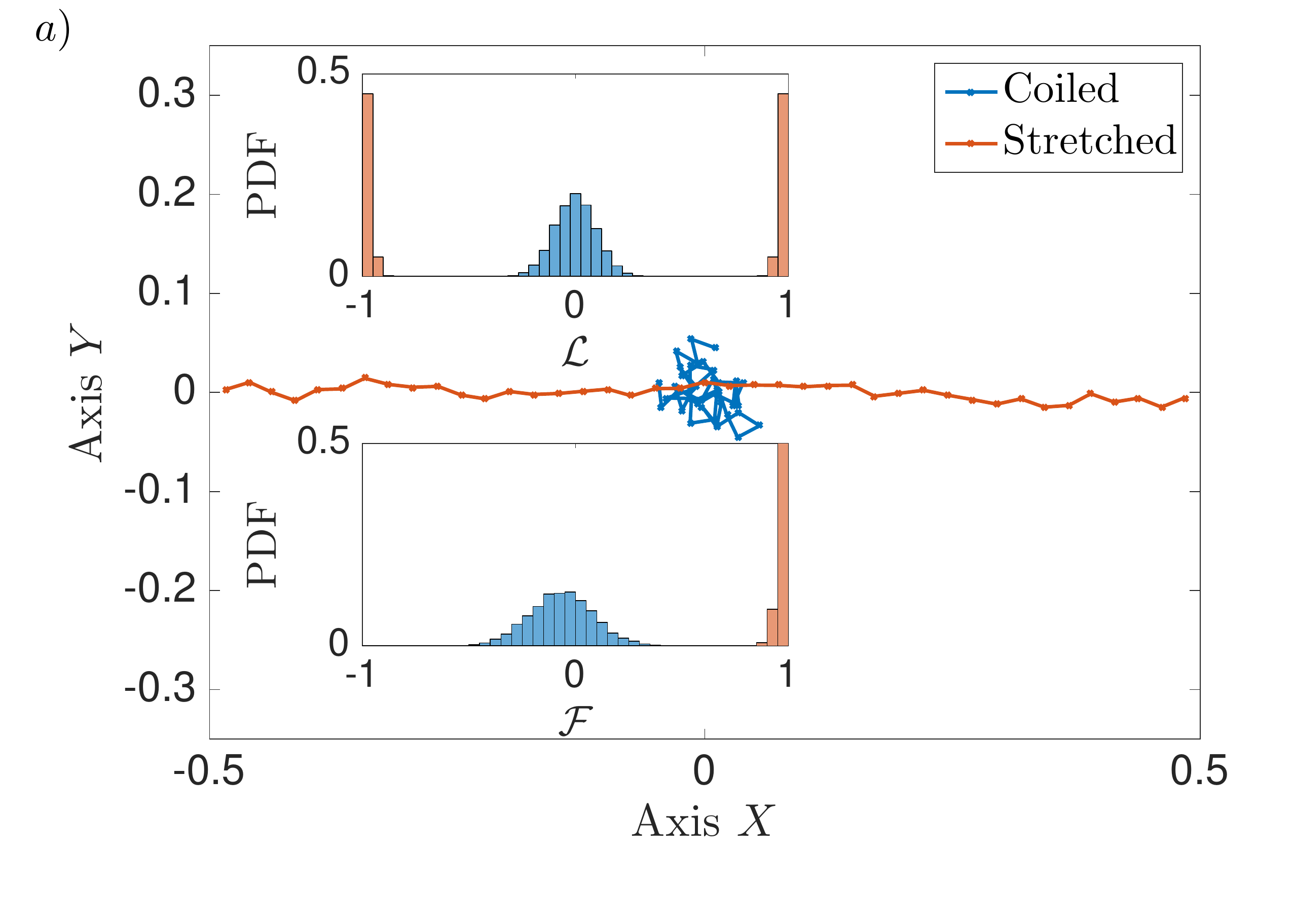}
\includegraphics[width=.48\textwidth]{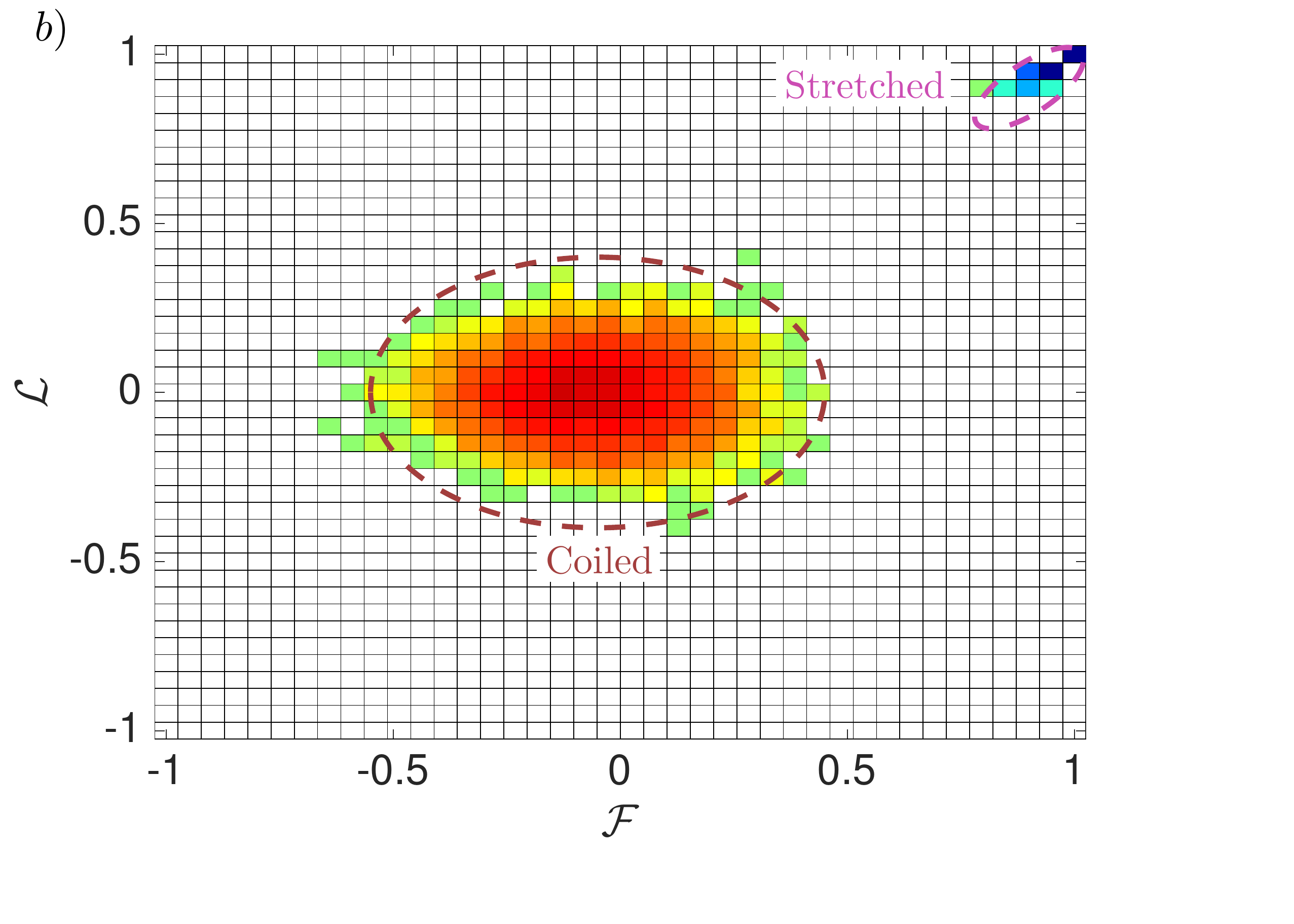}
\caption{(color online) Extension and orientation of an elongated
  chain ($N=40$) at two values of Peclet numbers showing the chain in
  both the coiled state ($\rm Pe=0.05$) and the stretched state
  ($\rm Pe=0.3$).  a) Snapshots of a fiber in the coiled and stretched
  states. The insets display the histograms of $\mathcal{L}$ and
  $\mathcal{F}$ for both states.  b)Map showing the probability of
  occurrence for each discrete state in the
  $(\mathcal{F},\mathcal{L})$ plane around the coiled (hot color) and
  stretched state (cold color). \label{fig:two_states_CS_FL} }
\end{figure*}

\begin{itemize}
\item A stretched configuration, where the fiber is oriented along the
  fluid streamline. This state corresponds to the case when
  $\rm Pe\gg1$, i.e.\ when fluid stretching prevails over thermal
  fluctuations. The probability density function (PDF) of $\mathcal{L}(t)$ then displays a two-peak
  shape (around $\pm 1$ depending on the orientation of the fiber with
  respect to the fluid stretching).  Meanwhile, the PDF of
  $\mathcal{F}(t)$ is peaked toward $1$ since all consecutive links
  are oriented in the same direction.
\item A coiled configuration, where the fiber is folded several times
  over itself. This state occurs when $\rm Pe\ll1$, i.e.\ when thermal
  fluctuations are predominant over fluid stretching.  In that case,
  the PDF of $\mathcal{L}(t)$ has a Gaussian distribution with a mean
  value equal to $0$ (coming from the random orientation of
  consecutive links). In the meantime, the PDF of $\mathcal{F}(t)$
  exhibit a Gaussian distribution with a mean value also equal to $0$
  (due again to the random orientation of consecutive links).
\end{itemize}
Another way to differentiate between coiled and stretched states is to
have a look at the fiber orientation and extension in the
$(\mathcal{F},\mathcal{L})$ plane (see
Fig.~\ref{fig:two_states_CS_FL}b). Fibers in the coiled state remain
around the configuration $(\mathcal{F},\mathcal{L}) \sim (0,0)$,
whereas fibers in the stretched state are close to
$(\mathcal{F},\mathcal{L}) \sim
(1,1)$. Figure~\ref{fig:two_states_CS_FL}b also provides additional
information on the nearby partially folded/unfolded states that fibers
explore. In particular, it appears that the range of nearby states
accessible depends on the relative importance of thermal fluctuations
to fluid stretching. This can be understood using intuitive arguments:
Folding occurs thanks to thermal fluctuations and thus higher noise
allows for more frequent and intense folding/unfolding events.

The Coil-Stretch (CS) transition is thus controlled by $\rm Pe$, i.e.\
by the balance between fluid stretching and thermal fluctuations. This
transition has been extensively studied in the literature since it is
key to understand the dynamics of elongated deformable particles in
flows (see, e.g.,
\cite{afonso2005nonlinear,ahmad2016polymer,ali2016semiflexible,
  beck2007ergodicity,de1974coil,gerashchenko2005single,kantsler2012fluctuations,
  plan2016tumbling,schroeder2003observation,schroeder2004effect,shaqfeh2005dynamics,
  watanabe2010coil,young2007stretch}). In the following, we analyze
the CS transition in the limit of very long particles with inertial
effects. For that purpose, their dynamics is assessed in terms of the
two observables ($\mathcal{L}$, $\mathcal{F}$) first in the overdamped
case to extract the key features before characterizing how inertia
affects these results.

\subsection{Overdamped case}
\label{sec:CST_Overdamped}
The transition from one state to another one is governed by the action
of the fluid on the fiber: Coiled fibers can be stretched when fluid
stretching is strong enough compared to thermal fluctuations, whereas
stretched fibers can become coiled if thermal fluctuations are locally
predominant. As a result, fibers orientation and extension evolve with
time, going through multiple states. The transition from one state to
another one is not instantaneous but occurs within a finite transition
time before relaxing to a steady state. This transition time depends
in a complex way on the Peclet number, the fiber length and the
initial state. To compare the temporal evolution of fibers, we have
characterized the fiber extension after a given time immersed in a
flow (here, we have retained a time roughly equal to $10$ times the
transition time). Fig~\ref{fig:avgL_fn_FL} displays the fiber
extension averaged over several realizations of the flow
$\langle\mathcal{|L|}(t)\rangle_{ens}$ taken after this time.
Observations that can be drawn from this plot are two-fold:
\begin{figure}[ht]
	\includegraphics[width=\columnwidth]{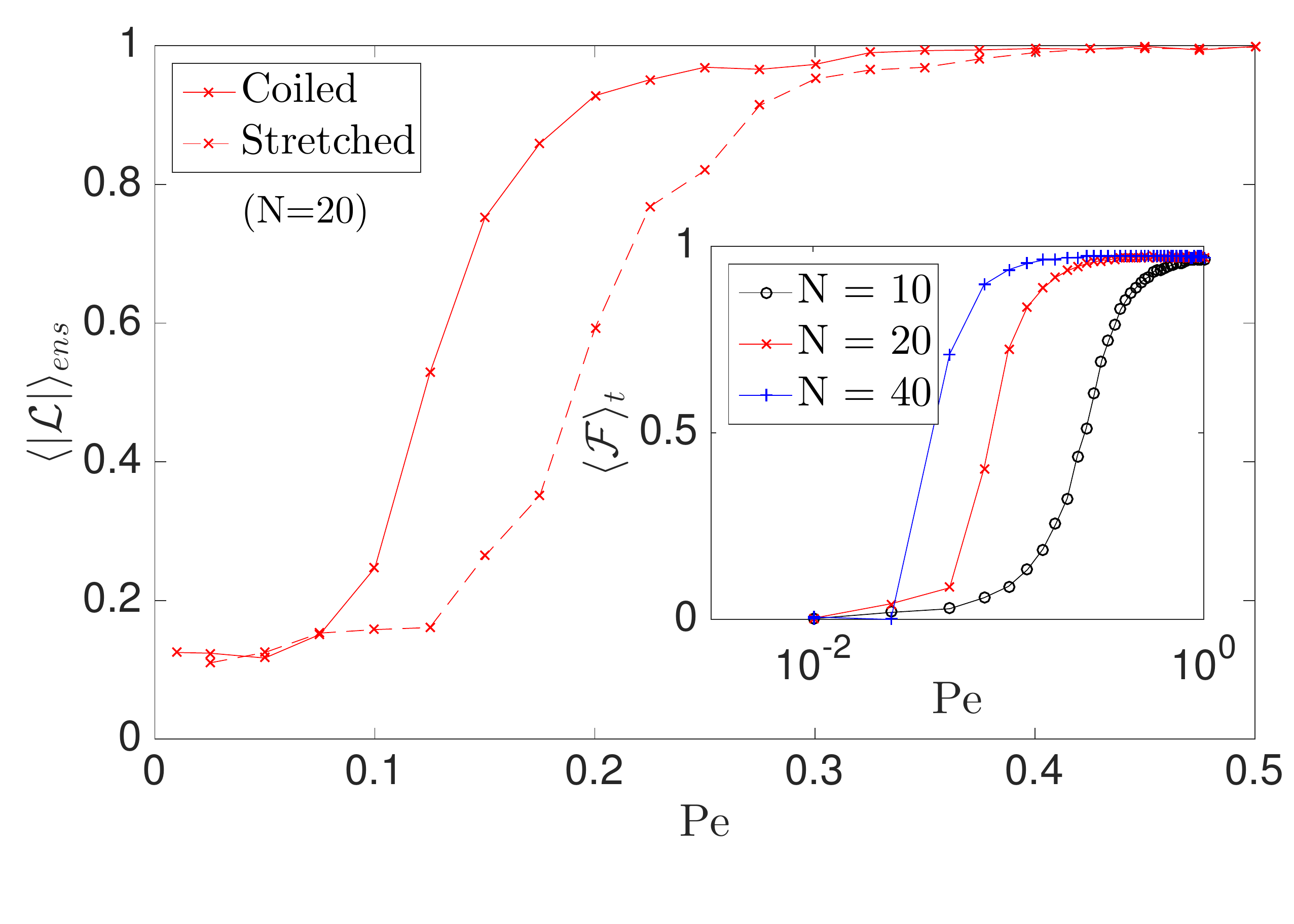}
	\caption{Fiber extension and orientation as a function of $\rm Pe$ 
		showing the sub-critical phase transition for 
		$N=10, 20, 40$. Sub-critical phase transition visible
		when plotting the average over several realizations of the
		flow $\langle\mathcal{|L|}\rangle_{ens}$. Inset showing
		the average orientation over time $\langle\mathcal{F}\rangle_{t}$.}
	\label{fig:avgL_fn_FL}
\end{figure}
\begin{itemize}
\item First, regardless of the initial configuration (stretched or
  coiled), $\langle \mathcal{|L|} \rangle_{ens}$ is characterized by a
  sudden increase when the Peclet number is increased. For long fibers
  (here $N\gtrsim10$), a fiber is indeed stretched when fluctuations
  are small ($\rm Pe\gtrsim1$) whereas it remains in a coiled state
  when fluctuations are large ($\rm Pe\ll1$). Besides, the averaged
  value of the fiber extension $\langle \mathcal{|L|} \rangle_{ens}$
  is proportional to $1/\sqrt{N}$ at small Peclet number. This can be
  explained using simple arguments: When thermal fluctuations are
  predominant over fluid stretching ($\rm Pe\ll1$), the orientation of
  each elementary link is independent of its neighbors. From the
  Central Limit Theorem, the fiber extension $\mathcal{L}$ is
  converging at large $N$ toward a Gaussian with a zero mean value. We
  thus have $\langle \mathcal{L} \rangle_{ens}=0$, while its modulus
  $\langle \mathcal{|L|} \rangle_{ens}$ is proportional to
  $1/\sqrt{N}$.
\item Second, there exists a conformation hysteresis for long fibers,
  meaning that the transition from the stretched state occurs at a
  Peclet number lower than the transition from the coiled state.
  There is thus a range of Peclet numbers where both states are
  concurrently stable to small fluctuations.  This hysteresis can be
  interpreted as resulting from higher internal constraints found in a
  stretched fiber. We have indeed seen in Sec.~\ref{sec:stat_state}
  that the tensions behave quadratically in a stretched configuration,
  while they are at best linear in a coiled fiber: this makes it
  harder to start folding a stretched fiber than to start unfolding a
  coiled fiber.  In the hysteresis band, both states coexist since
  fluctuations are high enough to partially fold stretched fibers and
  to partially unfold coiled ones. Therefore, a large spectrum of
  fiber orientations and extensions can be reached in this range of
  Peclet numbers. This hysteresis is very similar to the hysteresis
  loop that has been observed for long polymers or DNA molecules in a
  flow \cite{shaqfeh2005dynamics}.
\end{itemize}

To further assess the effect of $\rm Pe$ on $\mathcal{L}$ and
$\mathcal{F}$, the time-averaged fiber extension and orientation has
been extracted from numerical simulations when a stationary state is
reached. The inset in Fig.~\ref{fig:avgL_fn_FL} displays the
orientation $\langle \mathcal{F} \rangle$ as a function of the Peclet
number $\rm Pe$. It confirms that fibers are coiled for $\rm Pe\ll1$
and stretched for $\rm Pe\gtrsim1$.  It also appears that the
evolution of $\langle \mathcal{F} \rangle$ with $\rm Pe$ is monotonic:
It decreases toward $0$ when $\rm Pe$ decreases. The limit at
$\rm Pe\rightarrow0$ can be understood using simple handwaving
arguments: When thermal fluctuations prevails ($\rm Pe\ll1$), the
orientation of each segment is purely random and thus the
probabilities for consecutive links to be either aligned or reversed
is the same. As a result, the average fiber orientation
$\langle \mathcal{F} \rangle$ is equal to $0$. It should be noted here
that hysteresis is not visible since results have been averaged over
time, i.e. the dependence on the initial conditions has been lost in
the process.

Fiber orientation and extension are also affected by its length. It
appears from Fig.~\ref{fig:avgL_fn_FL} that the CS transition becomes
independent of the fiber length for sufficiently large values of
$N$. This can be be understood using qualitative arguments: We have
seen in Sec.~\ref{sec:stat_state_stretched} that the internal
constraints within a stretched fiber have a parabolic shape with a
maximum value in the middle of the fiber $\propto N^2$. As a result,
folding around the middle of the fiber becomes harder as the fiber
length $N$ increases but it remains possible to fold a long stretched
fiber close to its edges or close to already folded links (where the
internal constraint is smaller). Thus, when $N$ becomes large, the
Stretched-to-Coil transition occurs mostly through multiple folding of
the fiber around its edges or around folded links. The transition thus
occurs at similar values of the Peclet number as the
Coiled-to-Stretched transition, which is the exact opposite process
where multiple unfolding occurs. To further confirm these trends, the
fiber extension averaged over time $\langle\mathcal{|L|}\rangle_t$ has
been characterized as a function of both $N$ and $\rm Pe$.

\begin{figure}[ht]
  \includegraphics[width=\columnwidth]{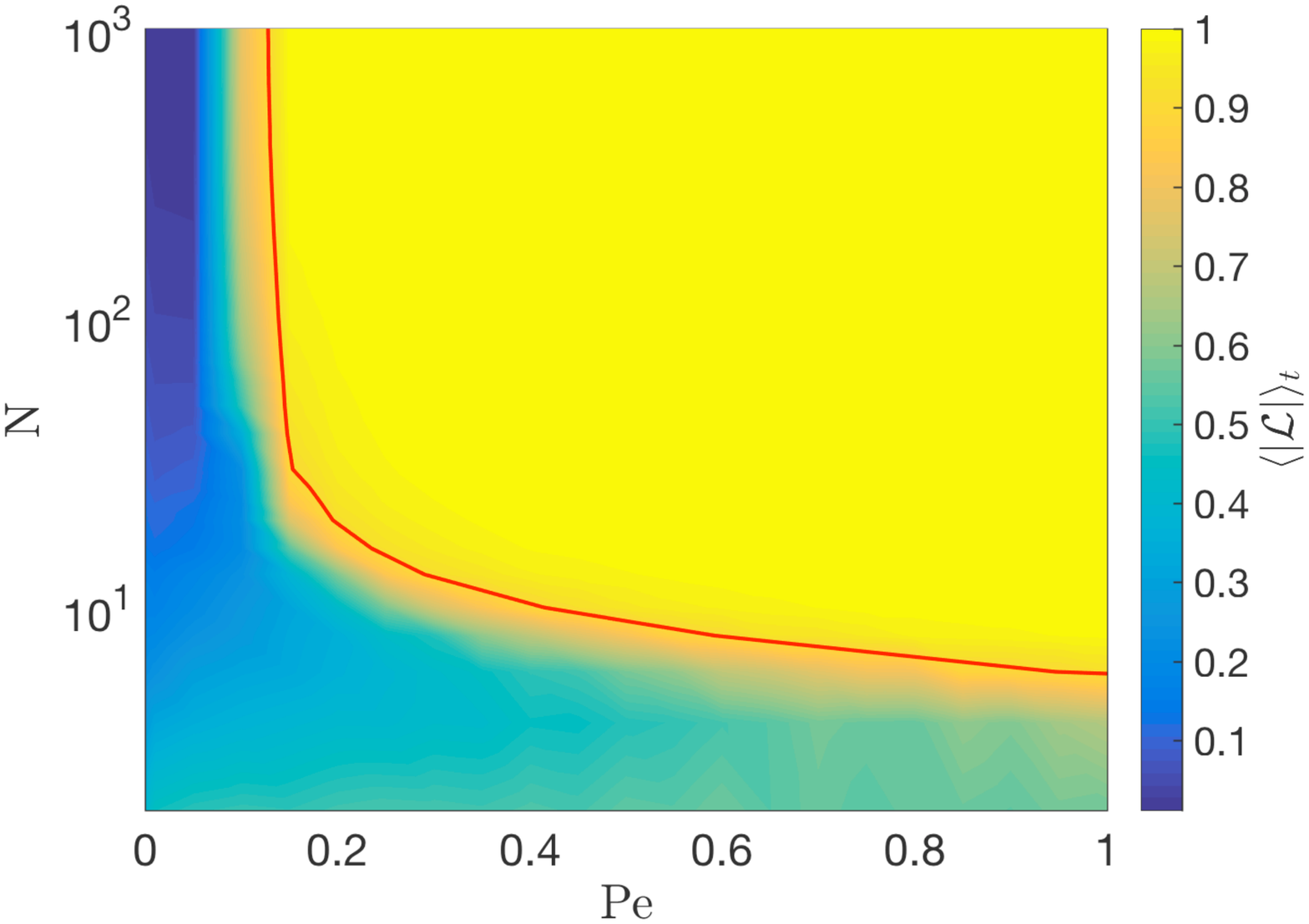}
  \caption{(color online) Average fiber extension over time
    $\langle \mathcal{L} \rangle_t$ in the $(\rm Pe,N)$ plane. The red
    solid line corresponds to the critical Peclet number
    $\rm Pe^\star$ at which the fiber extension exceeds a threshold
    $\langle\mathcal{|L|}\rangle_t>L^{\star} = 0.9$. This shows the
    existence of an asymptotic regime for sufficiently long fibers.}
  \label{fig:critical_line_Pe_N}
\end{figure} 
Results are plotted in Fig.~\ref{fig:critical_line_Pe_N}: One observes
that a long fiber (here $N>10$) is usually coiled for
$\rm Pe\lesssim0.1$ while it remains stretched for $\rm Pe\gtrsim0.4$.
The transition between the coiled and stretched states can be
distinguished using a critical Peclet number $\rm Pe^{\star}$, defined
as the value at which the averaged fiber extension exceeds a certain
threshold $\langle\mathcal{|L|}\rangle_t>L^{\star}$.  The solid red
line in Fig.~\ref{fig:critical_line_Pe_N} displays this critical
Peclet number for $L^{\star} = 0.9$. The CS transition has a
non-trivial dependence on the fiber length: It becomes independent of
the fiber length $N$ for sufficiently long fibers (here $N\gtrsim30$),
but it occurs at increasing Peclet numbers with small fibers (here
$30\gtrsim N\gtrsim4$).

\subsection{Inertial case}
\label{sec:CST_St}
We now characterize the effect of inertia on the coil/stretch
transition. Drawing on the observations made in the overdamped case,
we fix the fiber length to $N=20$ and characterize its orientation as
a function of both $\rm Pe$ and $\rm St$. The dynamics is impacted by
the inertia of each bead. In particular, two regimes can be identified
depending on whether the Stokes number is greater or higher than
$1/8$. Indeed, as it was shown in Sec.~\ref{sec:stat_state_stretched}
(in the absence of noise), for $\rm St < 1/8$ the eigenvalues of a
stretched chain are real ---\,see
Eq.~(\ref{eq:lower_bound_muM}). Therefore, at small Stokes numbers,
the inertial dynamics can be seen as the one of an overdamped chain in
a synthetic compressible flow where the effective compression rate
reads $\sigma(1+2\,{\rm St})$.

\begin{figure}[h]
	\includegraphics[width=\columnwidth]{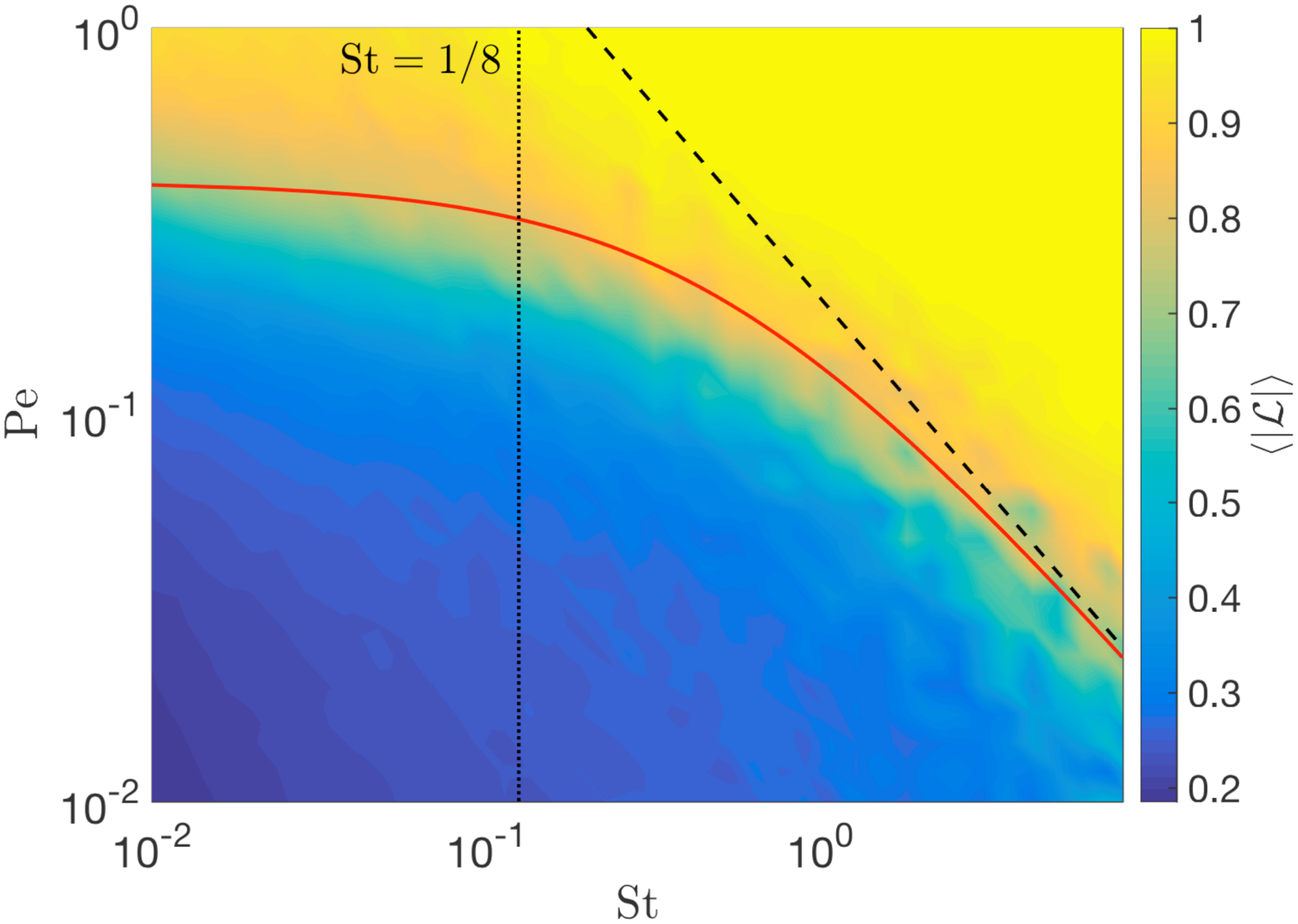}
	\caption{Time-averaged extension $\langle\mathcal{L}\rangle_t$
          in the $(\rm St,\rm Pe)$ plane.
          The dotted vertical line shows the critical value
          ${\rm St} =1/8$. The solid and dashed lines display the
          prediction
          $\rm Pe^\star(\rm St) = \rm Pe^\star(0)/(1+2\,\rm St)$ and
          the asymptotic behavior
          $\rm Pe^\star(\rm St)= Pe^\star(0)/(2\,\rm St)$,
          respectively, with ${\rm Pe}^\star(0)=0.4$.}
  \label{fig:critical_line_Pe_St}
\end{figure}
\begin{figure*}[tp]
  \includegraphics[width=.32\textwidth]{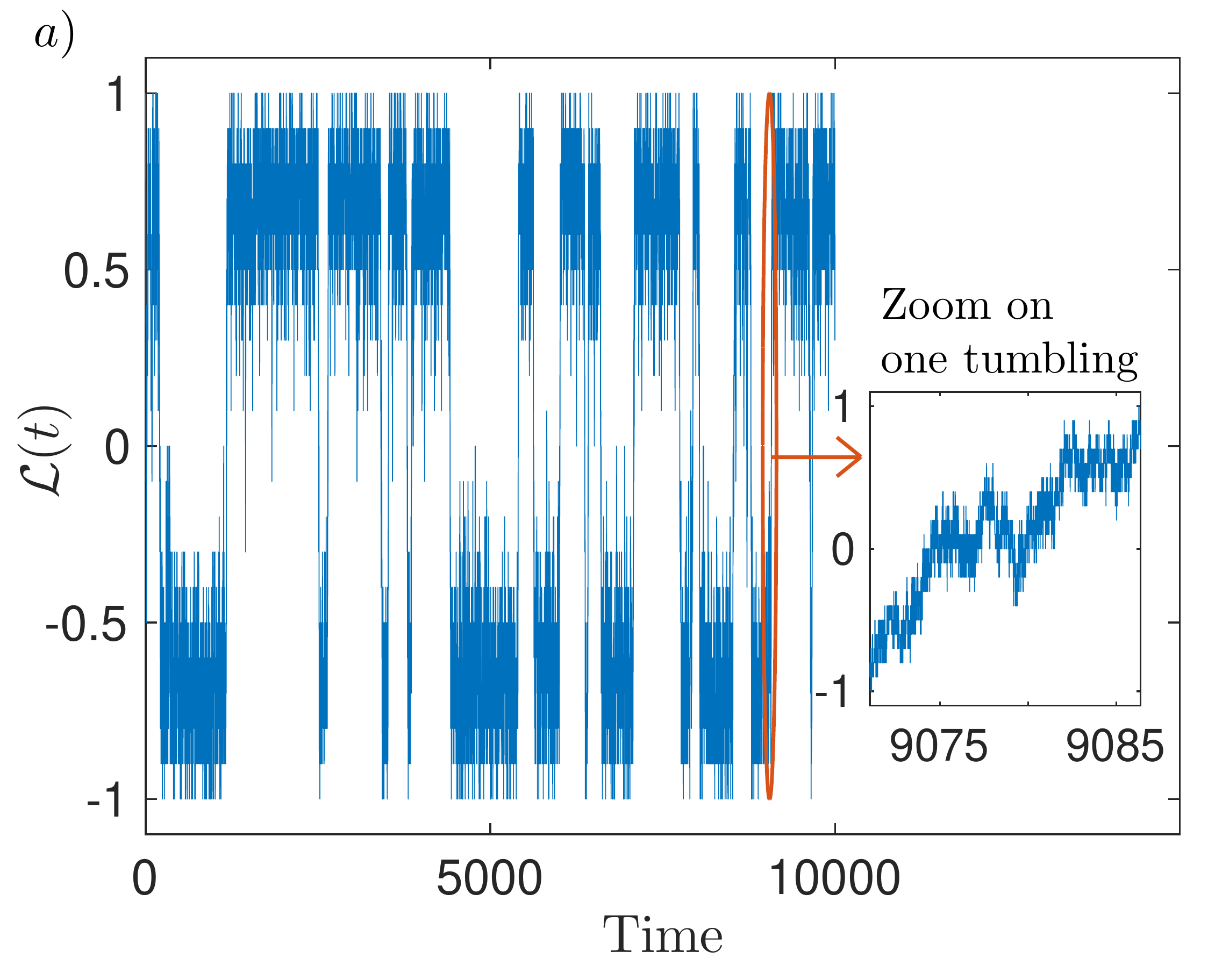}
  \includegraphics[width=.32\textwidth]{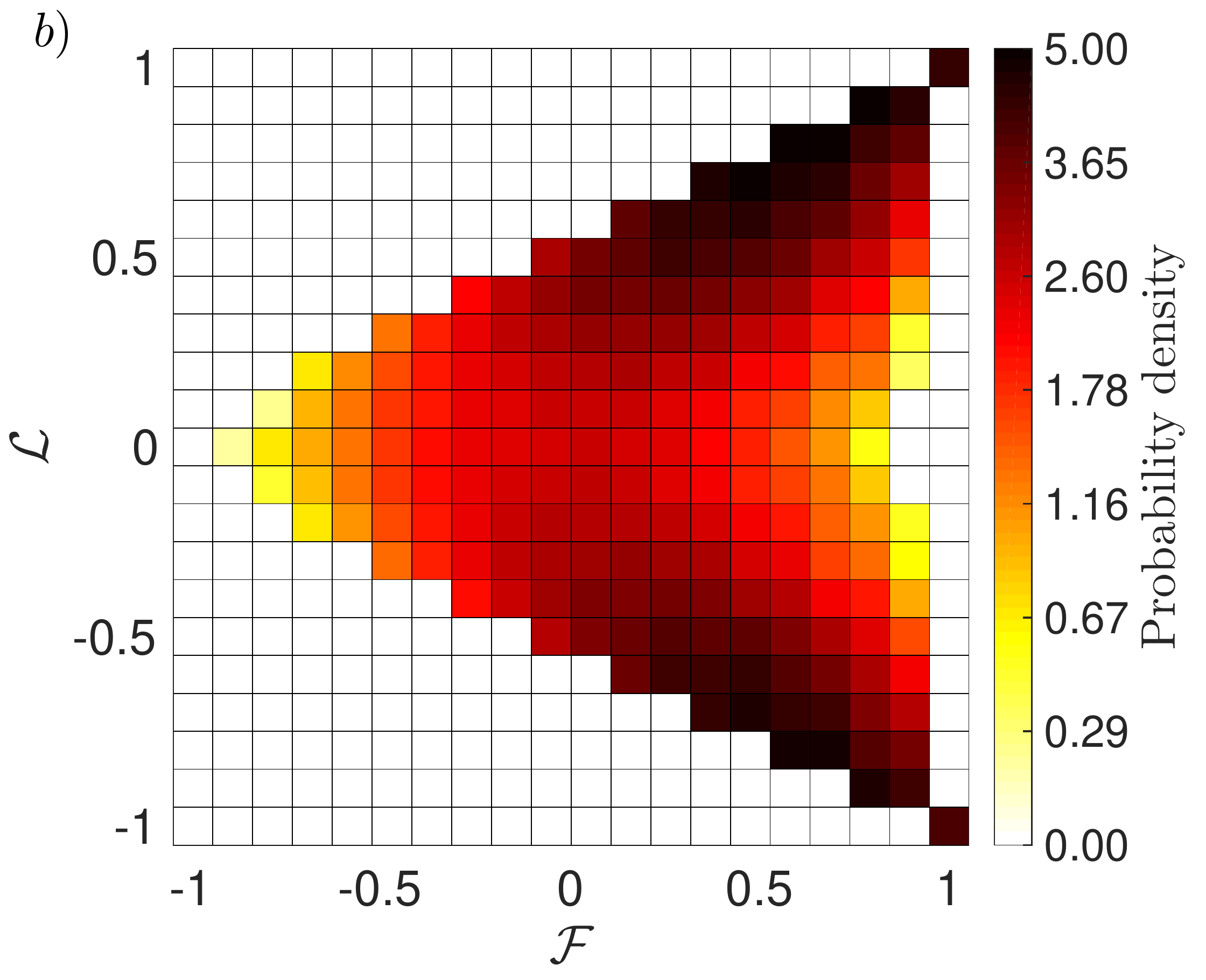}
  \includegraphics[width=.32\textwidth]{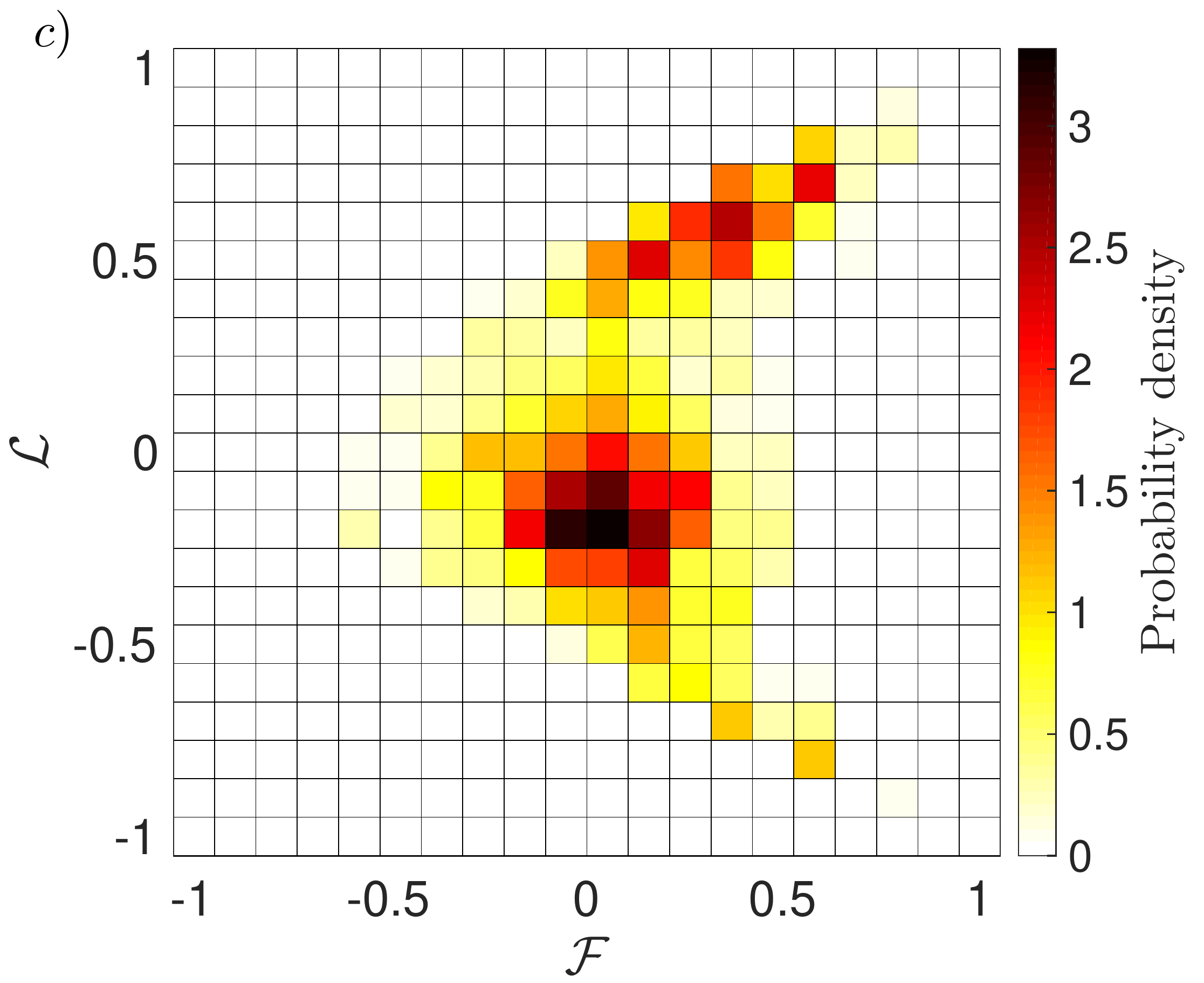}
  \caption{{Evolution of the fiber orientation for $N=20$ near the CS
      transition (here $\rm Pe=0.14$) in the overdamped case.  a)
      Fiber orientation $\mathcal{L}(t)$ as a function of time.  Inset
      showing a zoom on a selected tumbling event. b) Probability of
      occurrence for each discrete state over the whole simulation in
      the $(\mathcal{F},\mathcal{L})$ plane. c) Analogous to figure b
      but over a selected tumbling event.}}
	\label{fig:all_FL}
\end{figure*}
 Figure~\ref{fig:critical_line_Pe_St} shows the time-averaged fiber
extension $\langle\mathcal{|L|}\rangle_t$ in the $(\rm Pe,\rm St)$
plane. One clearly observes that for $St<1/8$, the CS transition
occurs at a critical value of the Peclet number that is compatible
with the formula $\rm Pe^\star(\rm St) = \rm Pe^\star(0)/(1+2\rm St)$
obtained using the effective compression rate described
above. Surprisingly, this behavior describes also what is happening
for $St>1/8$. At very large Stokes numbers, the critical Peclet number
decreases as a power-law $\rm Pe^\star(\rm St) \propto \rm
St^{-1}$. This can be interpreted using dimensional analysis as
follows: The CS transition results from the competition between
thermal fluctuations and fluid stretching and thus occurs when both
contributions are balanced. According to Eq.~(\ref{eq:evol_Xi}), this
means $1/ \rm St \sim 1/(\rm Pe\,\rm St^2)$ and thus
${\rm Pe} \sim {\rm St}^{-1}$.

\section{Tumbling}
 \label{sec:tumbling}

\subsection{Principle and mechanism}
\label{sec:tumbling_theory}

Another typical feature of fiber dynamics is the existence of tumbling
events. This is illustrated in Fig.~\ref{fig:all_FL}a: A fiber is
trapped in one of the stretched configuration with $\mathcal{L}=\pm 1$
for a long time until a sufficiently high fluctuation makes it tumble
toward the stretched configuration with a reversed orientation
$\mathcal{L}=\mp 1$.  The inset of Fig.~\ref{fig:all_FL}a shows a
focus on one of the tumbling events: It occurs due to a favorable
sequence of thermal fluctuations that allows the fiber to transition
from the stretched to the coiled state, before unfolding toward the
reversed stretched configuration. This process is thus very similar to
the tumbling-through-folding motion that has been recently
characterized for trumbbells \cite{plan2016tumbling}. It also
transpires from Fig.~\ref{fig:all_FL}a that, once the coiled state is
reached, tumbling does not necessarily occur since the fiber can
unfold back toward its original state. This is what happens near $t=5000$
where $\mathcal{L}$ departs from $-1$, reaches $0$, but then turns
back to its earliest negative orientation.

Further information on the intermediate states explored during the
tumbling event can be obtained by studying the probability of
occurrence for each discrete state in the $(\mathcal{F},\mathcal{L})$
plane. This is displayed in Figs.~\ref{fig:all_FL}b and
\ref{fig:all_FL}c that illustrate two features of tumbling events:
\begin{itemize}
\item \emph{The tumbling-through-folding concept:}\/ Fibers spend most
  of the time close to the stretched states where
  $(\mathcal{F}\approx1,\mathcal{L}\approx\pm1)$ and tumbling events
  appear to take place preferentially by going through a randomly
  coiled state $(\mathcal{F}\approx0,\mathcal{L}\approx0)$. This
  indicates that tumbling does not correspond to a rigid flip of the
  fiber.
\item \emph{The signature of intermediate states:}\/ The fibers seem
  to stay temporarily close to meta-stable states that where described
  in the stability analysis of Sec.~\ref{sec:stat_state_inter}. In the
  present case, the state where
  $(\mathcal{F}\approx 0.15,\mathcal{L}\approx-0.2)$ is rather
  frequent during the tumbling transition. It corresponds to the
  partially coiled configuration shown in
  Fig.~\ref{fig:eigval_diffstates}.
\end{itemize}

Tumbling is defined here as the change of the fiber extension
$\mathcal{L}$ from $\pm1$ to the opposite value. The persistence time
$\tau_{t}$ corresponds to the time spend by a fiber in an extended
state. In that sense, $\tau_{t}$ measures the time separating two
tumbling events. As for the CS transition, tumbling has been studied
previously in various flows \cite{einarsson2016tumbling,
  gustavsson2014tumbling,parsa2012rotation,plan2016tumbling,turitsyn2007polymer,
  vincenzi2013orientation}. We characterize here the tumbling dynamics
in the case of very long fibers in terms of the two observables
($\mathcal{L}$, $\mathcal{F}$) in the overdamped case first before
assessing the effect of inertia on tumbling.

\subsection{Overdamped case}
\label{sec:tumbling_Overdamped}

\begin{figure}[ht]
  \includegraphics[width=\columnwidth]{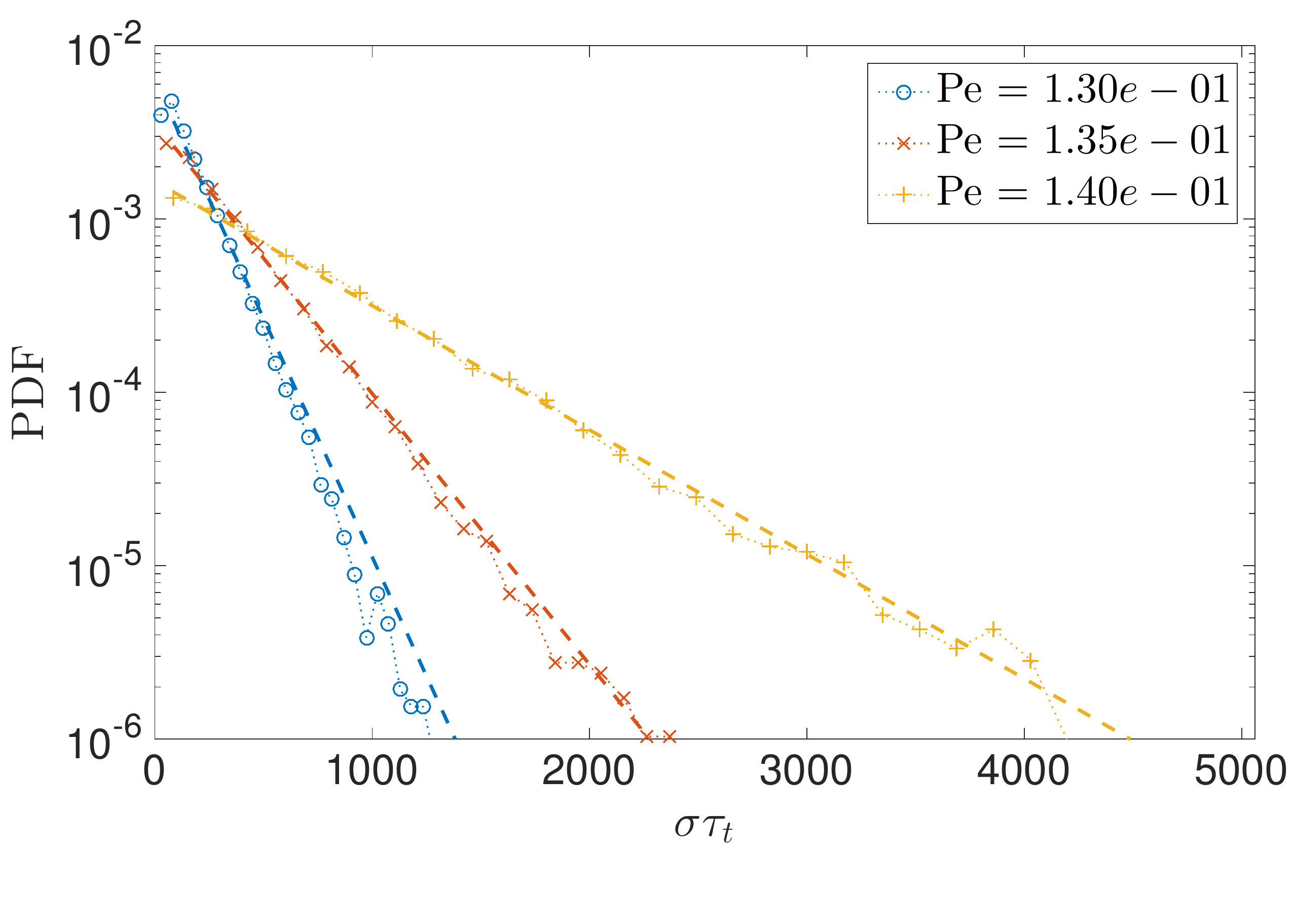}
  \caption{Probability density function (PDF) of the persistence time
    for $N=20$ in the overdamped case for various values of the Peclet
    number (slightly above the CS transition). The dotted lines
    correspond to the exponential law
    $p(\tau_{t})=\text{exp}(-\tau_{t}/\tau_{t}^{avg})/\tau_{t}^{avg}$.}
  \label{fig:PDF_Ttumble}
\end{figure}

As revealed by Fig.~\ref{fig:all_FL}.a, the persistence time is
distributed randomly. We have performed simulations with the same
fiber length (here $N=20$) over longer times to have sufficient
tumbling events to extract the PDF of $\tau_{t}$.  It is plotted in
Fig.~\ref{fig:PDF_Ttumble}) for various values of the Peclet number
(slightly above the critical Peclet $\rm Pe^\star$ at which the CS
transition occurs). First, it can be seen that the value of the 
persistence time is high, especially compared to recent experimental 
data on polymer dynamics in extensional flows which measured a 
residency time around $6s$ for a shear rate of $0.86s^{-1}$. 
\cite{perkins1997single}. Second,
the PDF turns out to have an exponential tail for
large $\tau_{t}$, i.e.\
$p(\tau_{t})\propto\text{exp}(-\tau_{t}/\tau_{t}^{avg})$ with
$\tau_{t}^{avg}$ the average persistence time. As for trumbbells in an
extensional flow \cite{plan2016tumbling}, this trend can be predicted
using the Freidlin--Wentzell large deviation theory
\cite{freidlin1998random} that characterizes the mean time required to
exit from a domain due to small perturbations. Regarding tumbling as a
fiber escaping from an attractor of a stochastic dynamical system in
the limit of small noise, the PDF of the exit time has an exponential
tail.

\begin{figure}[ht]
  \includegraphics[width=\columnwidth]{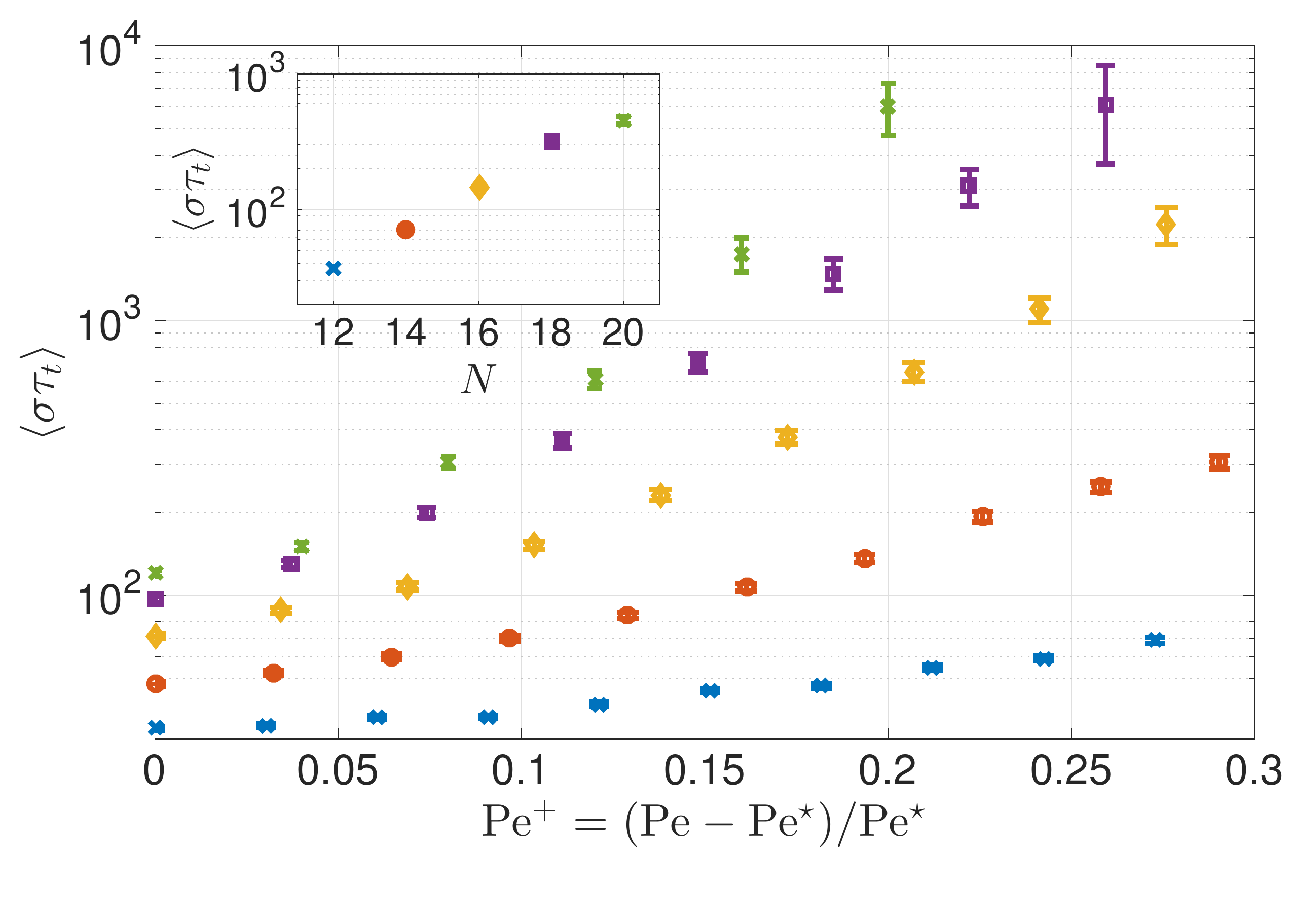}
  \caption{Average persistence time $\langle \tau_t\rangle$ as a
    function of the normalized Peclet number
    $\rm Pe^+=(\rm Pe-\rm Pe^{\star})/\rm Pe^{\star}$.  Inset showing
    the evolution of the persistence time $\langle \tau_t\rangle$ with
    the chain length $N$.}
  \label{fig:Ttumble_Pe_N}
\end{figure}

Furthermore, the theory also predicts that the mean exit time
increases exponentially when the amplitude of the noise
decreases. This relates to the fact that it becomes harder for fibers
to exit the stretched state when the amplitude of thermal fluctuations
decreases. This trend is confirmed in Fig.~\ref{fig:Ttumble_Pe_N},
where the average persistence time is plotted as a function of the
reduced Peclet number
$\rm Pe^+=(\rm Pe-\rm Pe^{\star})/\rm Pe^{\star}$. When fluctuations
are small enough (here for $\rm Pe^{+}>0.1$), one observes an
exponential increase of $\langle \tau_t\rangle$ with $\rm Pe^+$.  In
addition, the persistence time also increases very rapidly as a
function of the chain length $N$ (see the inset of
Fig.~\ref{fig:Ttumble_Pe_N}). This means that fibers are trapped in
either stretched states as $N\rightarrow\infty$ and that there is a
loss of ergodicity as the fiber length diverges at a fixed Peclet
number. A similar ergodicity breaking has been reported for the CS
transition \cite{beck2007ergodicity}: For a fixed noise amplitude
(Deborah number in the original article), the chains become
kinetically trapped either in the coiled or stretched states as their
length diverges. This has been explained theoretically by reducing the
problem to thermally activated transitions over an energy barrier
described by a rate theory. The same applies here: The energy needed
for a fiber to escape from one attractor (or state) to another is
proportional to the chain length, since the number of links is
higher. As a result, the transition rate decays exponentially with the
fiber length $N$. From a phenomenological point of view, this relates
to the fact that tumbling requires a favorable sequence of thermal
fluctuations to get out of the stretched state.

\subsection{Inertial case}
\label{sec:tumbling_St}
Drawing on the numerical results obtained in the overdamped case, we
now characterize the effect of inertia on the tumbling dynamics of
fibers. In line with the previous analysis of the CS transition
including inertial effects, we have chosen to fix the fiber length to
$N = 20$ and to assess how the tumbling dynamics evolves with both
Peclet and Stokes numbers.

\begin{figure}[ht]
  \includegraphics[width=\columnwidth]{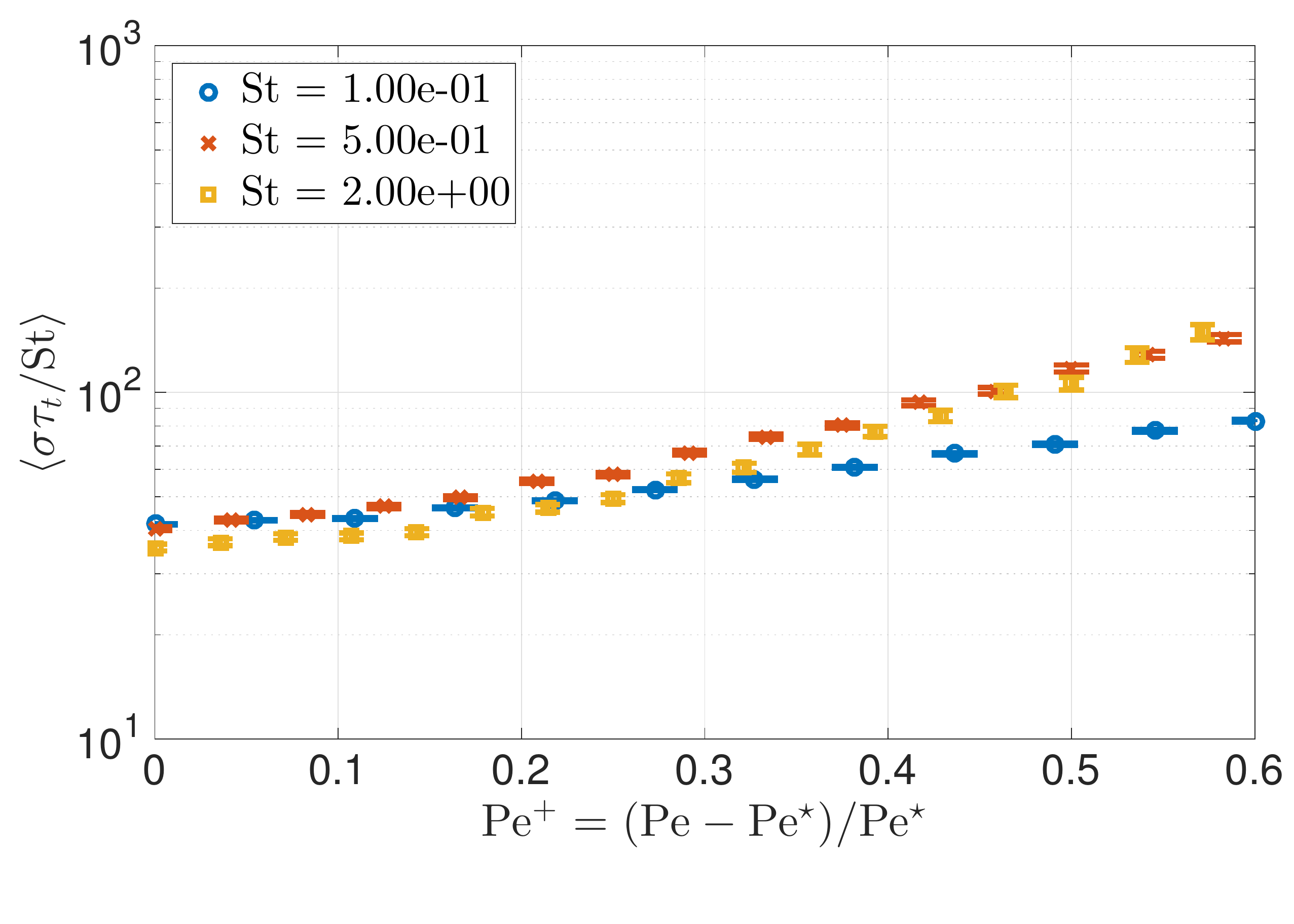}
  \caption{Average persistence time $\langle \tau_t\rangle$ as a
    function of the normalized Peclet number
    $\rm Pe^+=(\rm Pe-\rm Pe^{\star})/\rm Pe^{\star}$ for various
    Stokes number ($N = 20$).}
  \label{fig:Ttumble_St_Pe}
\end{figure}

The persistence time $\tau_t$ displays again an exponential tail at
large values (see the inset in Fig.~\ref{fig:Ttumble_St_Pe}). The mean
exit time is expected to increase exponentially as the amplitude of
the noise decreases for a given value of the Stokes number. This is
confirmed in Fig.~\ref{fig:Ttumble_St_Pe}) that displays the evolution
of the persistence time as a function of the reduced Peclet number
$Pe^+$ for three values of the Stokes number (resp. $0.1$, $0.5$ and
$2$). Besides, it also appears from Fig.~\ref{fig:Ttumble_St_Pe} that
the persistence time increases with the Stokes number and that two
regimes can be identified:
\begin{itemize}
\item Close to the CS transition ($\rm Pe^+\lesssim0.1$), the
  persistence time increases linearly with the Stokes number and all
  curves collapse on a single master curve when plotting
  $\tau_{t}/\rm St$ as a function of the reduced Peclet $\rm Pe^+$;
\item At larger Peclet numbers ($\rm Pe^+\gtrsim0.1$), different
  behaviors are displayed by the two families of
  particles. Low-inertia particles ($\rm St<1/8$) appear indeed to
  tumble at a higher rate than high-inertia particles ($\rm St>1/8$).
\end{itemize}

\section{Concluding remarks}

The dynamics of inertial deformable chains has been explored in the
case of an extensional flow. In particular, we have assessed the role
of the Peclet number, Stokes number and chain length on the
Coil-Stretch transition and tumbling phenomena. Numerical results have
confirmed the conformation hysteresis between coiled and stretched
states for sufficiently long fibers ($N\gtrsim10$). It has also been
seen that this transition depends non-linearly on these three
parameters: It becomes independent of the chain length $N$ for
sufficiently long fibers ($N\gtrsim 20-30$), while it evolves
proportionally to $1/(1+2\rm St)$ at a given fiber size. Similarly,
tumbling events have been confirmed in the case of an extensional flow
close to the Coil-Stretch transition. Numerical results support recent
simulations, which showed a loss of ergodicity as the chain length goes
to infinity (meaning that fibers are kinetically trapped in either
coiled or stretched states). The persistence time has also been shown
to increase exponentially with the chain length and Peclet number,
while it increases non-linearly with the Stokes number.

These promising results on the dynamics of inertial chains in an
extensional flow call for further refinements and developments. In
particular, it is worth assessing how 3D simulations affect these
results and to see if recent results showing a higher tendency for
trumbbells to remain in the stretched state in 3D cases than in 2D
cases are confirmed. The next step will be to investigate the role of
fiber flexibility on the coil-stretch transition and on tumbling
dynamics. Another question remains to be explored: What is happening
when fluctuations are triggered by the flow itself rather than by
noise. This issue will be investigated by coupling the dynamics of
such fibers with turbulent velocity gradients (coming directly from
direct numerical simulations). The role of fluctuations both in the
intensity of the velocity gradient and in its direction will be
explored. In the general context of fibers in turbulent flows, further
studies are needed to evaluate the effect of preferential sampling and
preferential concentration of such fibers in the near-wall region
especially in the case of highly elongated and deformable
fibers. These issues will be probed in future studies by coupling the
dynamics of such fibers with direct simulations of wall-bounded
turbulent flows.

\section{Acknowledgments}
This work has been supported by the French government, through the
Investments for the Future project UCA$^{\rm JEDI}$ managed by the
National Research Agency (ANR) with the reference number
ANR-15-IDEX-01.  The work of C.H. was supported by the PRESTIGE
Program (grant PRESTIGE-2017-1-0025) coordinated by Campus
France. Through this PRESTIGE program, this research has received
funding from the People Program (Marie Curie Actions) of the European
Union’s Seventh Framework Program (FP7/2007-2013) under REA grant
agreement n.\ PCOFUND-GA-2013-609102. C.H.\ acknowledges the support
of the EU COST Action MP1305 ``Flowing Matter''.

\appendix

\section{Appendix}

 \subsection{Appendix 1: implementation in the general case}
 
In the following, we focus on the numerical implementation of the
fiber equation of motion in an extensional flow. In that case, it 
is given by Eq.~\ref{eq:evol_Xi} which can be re-written as:
\begin{eqnarray}
 \frac{{\rm d} \bm \xi_{i}}{{\rm d} t} &=& {\bm V}_i\\
 \frac{{\rm d} \bm V_{i}}{{\rm d} t} &=& 
  -\frac{\zeta}{m}\left[\bm V_i -\bm\xi_i\cdot\bm\nabla\bm u\right] 
 +\sqrt{\frac{2k_{\rm
     B}T\, \zeta}{m^2\ell_{\rm K}^2}}\,(\bm\eta_i-\bm\eta_{i-1}) \nonumber \\
&& + 2\lambda_i\,\bm \xi_i - \lambda_{i+1}\,\bm \xi_{i+1}
  - \lambda_{i-1}\,\bm \xi_{i-1},
     \label{eq:evol_Xi_app}
\end{eqnarray}

This equation is solved using a simple first order Euler--Maruyama
method with temporal discretization (time step $\Delta t$):
\begin{eqnarray}
  \bm V_i(t+\Delta t) & = & \bm V_i(t) +\delta\bm V_i(t) \nonumber \\  
  \bm \xi_i(t+\Delta t) & = & \bm \xi_i(t) + \Delta t\,\bm V_i(t)
  \label{eq:evol_XiVi}
\end{eqnarray}
with the variation of the bead velocity given by:
\begin{eqnarray}
  \delta\bm V_i(t) & = & -\Delta t\,\frac{\zeta}{m}\left[\bm V_i(t) -\bm\xi_i(t)\cdot\bm\nabla\bm u\right] \nonumber \\
                   &&+ \sqrt{\Delta t}\,K_{Br}(\bm\gamma_i-\bm\gamma_{i-1}) -\Delta t\,
                      (\Delta_\xi \,\bm\lambda)_i 
\end{eqnarray}
with $K_{Br} = \sqrt{2k_{\rm B}T\,\zeta}/(m\,\ell_{\rm K})$ the
diffusion coefficient for Brownian motion, $\gamma_i$ taken from a
Gaussian distribution (with zero mean and a standard deviation equal
to $1$), $\bm\lambda$ denotes the $N$-dimensional vector
$(\lambda_1,\dots,\lambda_N)^\mathsf{T}$, and ${\bm \Delta}_{\xi}$ is
such that
$(\bm\Delta_\xi\,\bm\lambda)_{i} = - 2\lambda_i \bm\xi_i +
\lambda_{i-1} \bm\xi_{i-1} + \lambda_{i+1} \bm\xi_{i+1}$.
  
The tension forces acting on each rigid segment is obtained by
imposing a constant distance between consecutive beads
$\left|\bm\xi_i(t+\Delta t)\right|^2 = \left|\bm\xi_i(t)\right|^2 $
for all $1\le i\le N$.  Using Eq.~(\ref{eq:evol_XiVi}), this leads to
\begin{equation}
 2 \bm\xi_i(t)\cdotp \bm V_i(t) + \Delta t \left|\bm
   V_i(t)\right|^2 = 0.
 \label{eq:Lambda_app}
\end{equation}
By writing the above equation at time $t+\Delta t$, we obtain
\begin{eqnarray}
  2 \,  \Delta t\,\left|\bm V_i(t)\right|^2 + 2 \,
  \bm\xi_i(t)\cdotp\delta\bm V_i(t) && \nonumber\\+ 4 \,
  \Delta t \,\bm V_i(t)\cdotp \delta\bm V_i(t) + \Delta
  t \,\left|\delta\bm V_i(t) \right|^2 &&=0 
 \label{eq:Lambda_app_2}
\end{eqnarray}
The above non-linearities does not allow for writing an explicit
solution.  Yet, one can note that the terms on the left-hand side
involve various powers of the time step $\Delta t$. For that reason,
we chose to decompose the tensions $\lambda_i$ as series in powers of
$\sqrt{\Delta t}$, i.e.
\begin{equation}
  \Delta t\,\bm\lambda = \sum_{k=1}^{\infty} \bm\lambda^{(k)} \,\Delta t^{k/2}.
 \label{eq:Lambda_decomp}
\end{equation}
The series starts with terms $\mathcal{O}(\sqrt{\Delta t})$ to account
for the tension that balances the noise.  We can then identify each
contribution to obtain a set of equations for each term:
\begin{itemize}
\item Terms in $\Delta t^{1/2}$:
  $$  \bm\xi_i\cdotp (\bm\Delta_\xi\bm\lambda^{(1)})_i = 
  K_{Br}\,\bm\xi_i\cdotp(\bm\gamma_i-\bm\gamma_{i-1}).
  $$
\item Terms in $\Delta t$
  $$
    \bm\xi_i\cdotp (\bm\Delta_\xi\bm\lambda^{(2)})_i = \left|\bm
      V_i\right|^2 - \frac{\zeta}{m}\bm\xi_i\cdotp \left(\bm V_i
      -\bm\xi_i\cdot\bm\nabla\bm u\right).
    $$
\item Terms in $\Delta t^{3/2}$
  $$    \bm\xi_i\cdotp (\bm\Delta_\xi\bm\lambda^{(3)})_i = 2\bm
      V_i\cdotp\left[
        K_{Br}\,(\bm\gamma_i-\bm\gamma_{i-1})-(\bm\Delta_\xi\bm\lambda^{(1)})_i\right].
  $$
\item Terms in $\Delta t^{2}$
  \begin{eqnarray}    \bm\xi_i\cdotp (\bm\Delta_\xi\bm\lambda^{(4)})_i &=& 2\bm
      V_i\cdotp\left[- \frac{\zeta}{m}\bm\xi_i\cdotp \left(\bm V_i
      -\bm\xi_i\cdot\bm\nabla\bm u\right) -
                                                                           (\bm\Delta_\xi\bm\lambda^{(2)})_i\right]
                                                                           \nonumber\\
&& + \frac{1}{2}\left|
        K_{Br}\,(\bm\gamma_i-\bm\gamma_{i-1})-(\bm\Delta_\xi\bm\lambda^{(1)})_i\right|^2.
  \nonumber
\end{eqnarray}
\end{itemize}
The above expansion can be continued to reach an arbitrary
precision. In practice, we stop at a given order and use the
corresponding approximation of the tensions to update the fiber
velocity. In the paper, we used the expansion to order $\Delta t$. The
terms order $\Delta t^{3/2}$ are stochastic with a zero mean, so that
the error is in average $\mathcal{O}(\Delta t^2)$.

\subsection{Appendix 2: implementation for small fibers}
In the case of fibers composed of beads that act as tracers in the
flow, the equation of motion simplifies to:
\begin{eqnarray}
 \frac{{\rm d} \bm\xi_{i}}{{\rm d} t} & = & \bm\xi_i\cdot\bm\nabla\bm u +  
 \sqrt{\frac{2k_{\rm B}T}{\ell_{\rm K}^2\,\zeta}}\,(\bm\eta_i-\bm\eta_{i-1})\nonumber\\
 && + 2\lambda_i'\,\bm \xi_i - \lambda_{i+1}'\,\bm \xi_{i+1}
 - \lambda_{i-1}'\,\bm \xi_{i-1},
     \label{eq:evol_Xi_tracers_app}
\end{eqnarray}
with $\lambda' = \lambda\,{m}/{\zeta}$.  This equation is solved using
a first order Euler--Maruyama method with time step $\Delta t$:
\begin{eqnarray}
  \bm \xi_i(t+\Delta t) & = & \bm \xi_i(t) + \delta\bm\xi_i(t)
\end{eqnarray}
with the variation of the bead position $\delta\bm\xi_i(t)$ given by:
\begin{eqnarray}
  \delta\bm\xi_i(t) = && \Delta t\, \bm \xi_i(t)\cdot\bm\nabla\bm u + \sqrt{\Delta t}\,K_{Br}'(\bm\gamma_i-\bm\gamma_{i-1}) \nonumber \\ 
  && -\Delta t\,
                      (\Delta_\xi \,\bm\lambda')_i 
\end{eqnarray}
with $K_{Br}' = \sqrt{(2k_{\rm B}T)/\,(\zeta\,\ell_{\rm K}^2)}$
the diffusion coefficient for Brownian motion, the $\gamma_i$'s taken
from a Gaussian distribution (with zero mean and a standard deviation
$1$).

The tension forces acting on each rigid segment are obtained by
imposing a constant distance between consecutive beads, i.e.\
$\left|\bm\xi_i(t+\Delta t)\right|^2 =  \left|\bm\xi_i(t) \right|^2$, leading to
\begin{equation}
  2 \bm\xi_i(t)\cdotp \delta\bm\xi_i(t) +
  \left|\delta\bm\xi_i(t)\right|^2 = 0
 \label{eq:Lambda_tracers_app}
\end{equation}
As in the inertial case, we decompose the tensions $\lambda_i'$ as 
\begin{equation}
  \Delta t\,\bm\lambda' = \sum_{k=1}^{\infty} \bm\lambda'^{(k)} \,\Delta t^{k/2},
 \label{eq:Lambda_decomp_ap}
\end{equation}
and identify contributions of different orders in $\Delta t^{1/2}$:
\begin{itemize}
 \item Terms in $\Delta t^{1/2}$:
  \begin{eqnarray}
   \bm\xi_i \cdotp (\Delta_\xi \,\bm\lambda'^{(1)})_i =  K_{Br}' \, \bm\xi_i \cdotp (\bm\gamma_i-\bm\gamma_{i-1}). \nonumber
   \label{eq:Lambda_tracer_1_app}
  \end{eqnarray}
 \item Terms in $\Delta t$:
 \begin{eqnarray}
   \nonumber \bm\xi_i && \cdotp (\Delta_\xi \,\bm\lambda'^{(2)})_i =  
   \bm\xi_i\cdotp\left(\bm\xi_i\cdotp\bm\nabla\bm u\right) - K_{Br}'(\bm\gamma_i-\bm\gamma_{i-1})\cdotp (\Delta_\xi \,\bm\lambda'^{(1)})_i \\ \nonumber
   && + \frac{1}{2}\left|K_{Br}'(\bm\gamma_i-\bm\gamma_{i-1})\right|^2 + \frac{1}{2}\left|(\Delta_\xi \,\bm\lambda'^{(1)})_i\right|^2. \nonumber
   \label{eq:Lambda_tracer_2_app}
  \end{eqnarray}
 \item Terms in $\Delta t^{3/2}$
 \begin{eqnarray}
   \nonumber \bm\xi_i&& \cdotp (\Delta_\xi \,\bm\lambda'^{(3)})_i =
   \left(\bm\xi_i\cdotp\bm\nabla\bm u-(\Delta_\xi \,\bm\lambda'^{(2)})_i\right)  \\
   && \, \cdotp \left(K_{Br}'(\bm\gamma_i-\bm\gamma_{i-1})-(\Delta_\xi \,\bm\lambda'^{(1)})_i\right). \nonumber
   \label{eq:Lambda_tracer_3_app}
  \end{eqnarray}
 \item Terms in $\Delta t^{2}$
 \begin{eqnarray}
   \nonumber \bm\xi_i && \cdotp (\Delta_\xi \,\bm\lambda'^{(4)})_i =  
   - \left(\bm\xi_i\cdotp\bm\nabla\bm u\right)\cdotp (\Delta_\xi \,\bm\lambda'^{(2)})_i\\ \nonumber
   &&\left((\Delta_\xi \,\bm\lambda'^{(1)})_i-K_{Br}'(\bm\gamma_i-\bm\gamma_{i-1})\right)\cdotp (\Delta_\xi \,\bm\lambda'^{(3)})_i\\ \nonumber
   && + \frac{1}{2}\left|\bm\xi_i\cdotp\bm\nabla\bm u\right|^2 +
      \frac{1}{2}\left|(\Delta_\xi \,\bm\lambda'^{(2)})_i\right|^2. \nonumber
   \label{eq:Lambda_tracer_4_app}
  \end{eqnarray}
\end{itemize}
The above approximation, with terms up to those of order $\Delta t^2$,
is used in the paper.

\bibliographystyle{apsrev4-1}
\bibliography{biblio_fibers}

\end{document}